\documentclass[12pt,a4paper]{article}

\usepackage{setspace}
\usepackage{authblk}

\usepackage{float}
\restylefloat{table}

\usepackage{graphicx}	 			
\usepackage{amsmath}
\usepackage{amssymb}
\usepackage{mathrsfs}
\usepackage{amsthm}
\usepackage{ulem}
\usepackage{graphicx}	

\usepackage{cancel}

\usepackage{natbib}
\usepackage{epsfig,amstext,amsfonts,color}

\usepackage{latexsym}

\newcommand{\E}{\operatorname{E}}
\newcommand{\Cov}{\operatorname{Cov}}
\newcommand{\Var}{\operatorname{Var}}

\newcommand{\bbeta}{\mathbf{\beta}}

\newcommand{\bmu}{\mathbf{\mu}}
\newcommand{\bpi}{\mathbf{\pi}}

\newcommand{\bz}{\mathbf{z}}
\newcommand{\bx}{\mathbf{x}}

\newcommand{\br}{\mathbf{r}}

\newcommand{\bt}{\mathbf{t}}
\newcommand{\bu}{\mathbf{u}}
\newcommand{\bv}{\mathbf{v}}

\newcommand{\bb}{\mathbf{b}}

\newcommand{\bd}{\mathbf{d}}

\newcommand{\bR}{\mathbf{R}}
\newcommand{\bA}{\mathbf{A}}

\newcommand{\bV}{\mathbf{V}}
\newcommand{\bQ}{\mathbf{Q}}

\newcommand{\bI}{\mathbf{I}}

\newcommand{\bW}{\mathbf{W}}

\newcommand{\bSigma}{\mathbf{\Sigma}}
\renewcommand{\mathbf}{\boldsymbol}

\newcommand{\bP}{\mathbf{P}}

\title{Score matching for compositional distributions}

\author[1,*]{Janice L. Scealy}
\author[1]{Andrew T. A. Wood}
\affil[1]{Research School of Finance, Actuarial Studies and Statistics, Australian National University, Canberra ACT 2601, Australia}
\affil[*]{Corresponding author's email: janice.scealy@anu.edu.au.~~~~~~~~~~~~~~~~~~~~~~~~~~~~~~~~}

\date{}

\makeatletter
\renewcommand\@biblabel[1]{}
\makeatother

\begin{document}

\maketitle

\begin{abstract}

Compositional data and multivariate count data with known totals are challenging to analyse due to the non-negativity and sum-to-one constraints on the sample space. It is often the case that many of the compositional components are highly right-skewed, with large numbers of zeros. A major limitation of currently available estimators for compositional models is that they either cannot handle many zeros in the data or are not computationally feasible in moderate to high dimensions. We derive a new set of novel score matching estimators applicable to distributions on a Riemannian manifold with boundary, of which the standard simplex is a special case. The score matching method is applied to estimate the parameters in a new flexible truncation model for compositional data and we show that the estimators are scalable and available in closed form. 
Through extensive simulation studies, the scoring methodology is demonstrated to work well for estimating the parameters in the new truncation model and also for the Dirichlet distribution.  We apply the new model and estimators to real microbiome compositional data and show that the model provides a good fit to the data.

\vspace{3.mm}
\end{abstract}

\section{Introduction}

Modern scientific measuring systems commonly record compositional data, which are $p$-dimensional vectors $\bu$ defined on the unit simplex 
\begin{equation*}
\Delta^{p-1}=\left\{(u_1,u_2,\ldots,u_p)^{\top} \in \mathbb{R}^p : \quad u_i \geq 0,\quad \sum_{i=1}^pu_i=1  \right\}.
\end{equation*}
A discrete form of compositional data can also arise through observing non-negative counts $\bx=(x_1,x_2,\ldots, x_p)^{\top}$ in $p$ categories with the constraint $m=\sum_{i=1}^px_i$, where $m > 0$ is a fixed integer constant which is known. Estimated proportions on the simplex can then be calculated using $\bx/m$.  Currently, there is no frequentist parametric analysis method available for these types of multivariate data which simultaneously satisfies all three of the following criteria: (i) fully flexible model; (ii) handles large numbers of zeros in the data; and (iii) tractable estimation for substantial $p$ (and sample size). In this article we define a new set of computationally fast estimators for both discrete and continuous compositional data models which satisfy these criteria.

Various different distributions have been proposed for modelling compositional data which take the constraints within $\Delta^{p-1}$ into account. In summary these are the Dirichlet distribution and its generalisations (e.g. Krzysztofowicz and Reese, 1993; Ongaro et al., 2020), the logistic normal distribution (Aitchison, 1986) and the folded Kent distribution based on the square root transformation (Scealy and Welsh, 2011;2014). Butler and Glasbey (2008) projected a multivariate Gaussian distribution onto the simplex and Leininger et al. (2013) applied a different projection (after truncation at zero) of a multivariate Gaussian distribution onto the simplex. More recently Tsagris and Stewart (2020) projected and folded a power-transformed multivariate Gaussian distribution onto the simplex.  

Maximum likelihood estimation for the projected, truncated and folded distribution parameters is not straight-forward due to the complexities of the densities and when $p$ is not small these methods are generally not tractable. Maximum likelihood estimation for the Dirichlet, logistic normal distribution and the Tsagris and Stewart (2020) model cannot be applied whenever there are zeros in the data since the log-likelihood is not finite. Stewart and Field (2010)  and Bear and Billheimer (2016) both modified the logistic normal model to handle zero's by separating the components each into two parts and modelling the zeros separately.  However, these two methods based on maximum likelihood estimation were applied only in the case where the dimension was small and Bear and Billheimer (2016) assumed that one of the components was always non-zero.  In the case of the Dirichlet distribution with some zero components, moment estimators are still valid, however this model is not fully flexible because it has a restrictive correlation structure, for example all covariances are negative between the components.

Multivariate count data with known totals are often modelled by using the multinomial distribution. However, as with the Dirichlet distribution, the covariance structure in the multinomial model is rather restrictive. One way to obtain a more flexible model is to introduce a set of latent variables to model the probability vector in the multinomial distribution. Both the Dirichlet distribution and the logistic normal distribution have been used for this purpose in the literature and these models are referred to as the Dirichlet multinomial model (e.g. Li, 2015) and the logistic normal multinomial model (e.g. Zhang and Lin, 2019). The logistic normal multinomial model is fully flexible and allows for general correlation structures, but estimation is rather difficult due to the fact that the marginal loglikelihood is not tractable. 

As an alternative to maximum likelihood estimation, Hyvarinen (2005) developed a set of score matching estimators for densities on the Euclidean space. A key advantage of this estimator is that it completely avoids having to calculate the normalising constant in the model.  In its original form, score matching assumed that the probability density function is differentiable over the entire space $\Bbb{R}^{p}$.  Hyvarinen (2007) extended score matching to densities defined on $\Bbb{R}^{p}_+$ or $\{ \Bbb{R}_+ \cup \{  0 \} \}^p$ by introducing weights into the objective function that are zero on the boundary of the sample space.  Yu et al. (2019) extended this approach to allow non-smooth weights in order to improve estimation efficiency. Liu et al. (2020) investigated score matching on Euclidean spaces with more complicated truncation boundaries and applied a weight function defined by the distance from a point in the domain to the boundary of the domain.  Mardia et al. (2016) and Mardia (2018) extended the score matching method to handle densities on a Riemannian manifold and Takasu et al. (2018) proposed various different scoring rules for spheres. 

In this article we define a new flexible model for continuous compositional data based on truncated distributions. To estimate the model parameters, we develop score matching estimators on a manifold with boundary. We show that these new estimators are consistent and efficient and are very simple to compute in closed form for the new truncated model for compositional data. This constitutes a major computational and methodological advance. The new models are motivated from analysing real microbiome count data with many zeros from Martin et al. (2018). For this data, the values of the count totals, $m$, are large compared with the sample size and it is possible to model $\bu=\bx/m$ as approximately continuous for estimation purposes. We also show that score matching can be used for consistent estimation in multinomial models with compositional latent variables, again with closed form solutions.

The remainder of the article is organised as follows. The new models for compositional data are introduced in Section 2 and score matching estimators are derived in Section 3. Section 4 contains  the results
of a simulation study. 
In Section 5, we analyse real microbiome count data and the article concludes with some discussion
in Section 6.

\section{Models}

\subsection{Continuous case: a general pairwise interaction model}

Assume that we have an iid (independent  and identically distributed) sample $\bu_i=(u_{i1},u_{i2}, \ldots, u_{ip})^{\top} \in \Delta^{p-1}$, $i=1,2,\ldots, n$, where $n$ is the sample size. Each compositional data vector $\bu_i$ is assumed to have the following density
\begin{equation}
\frac{1}{c_1(\bA,\bbeta)}\left( \prod_{j=1}^p u_j^{\beta_j} \right) \exp{\left( \bu^{\top} \bA \bu  \right)}, \quad \bu =(u_1,u_2,\ldots, u_p)^{\top} \in \Delta^{p-1},
\label{uqdensity}
\end{equation}
where $c_1(\bA,\bbeta)$ is the normalising constant and $\bA$ is a general $p\times p$ symmetric matrix and we assume that one of the eigenvalues  of  $\bA$ is zero due to the constraint that the compositional components sum to one.  Note that the matrix $\bA$ does not necessarily need to be negative semi-definite due to the fact that the simplex is a bounded space. The vector $\bbeta=(\beta_1,\beta_2,\ldots,\beta_p)^{\top}$ consists of shape parameters which satisfy $\beta_j > -1$ for $j=1,2,\ldots, p$. This model is permutation invariant and is very flexible. The parameter $\bbeta$ helps to control the shape of the distribution close to the zero boundary and $\bA$ helps to control the covariance and location of the data. If we set $\bbeta=\boldsymbol{0}$, then the density has the same number of parameters as the $(p-1)-$dimensional Gaussian distribution.

An equivalent density, whose parameterisation is more convenient, is
\begin{equation}
\frac{1}{c_2(\bA^*,\bb,\bbeta)} \left( \prod_{j=1}^p u_j^{\beta_j} \right) \exp{\left( \bu^{\top} \bA^* \bu + \bb^{\top} \bu   \right)}, \quad \bu =(u_1,u_2,\ldots, u_p)^{\top} \in \Delta^{p-1},
\label{uqdensity2}
\end{equation}
where $\bA^*$ is a $p\times p$ symmetric matrix  with $i,j$th element denoted by $a_{ij}$ and $\bb$ is a $p \times 1$ vector with $j$th element denoted by $b_j$. To account for the constraint and for identifiability we set the last row and column of $\bA^*$ to zero, that is $a_{1p}=a_{2p}=\ldots =a_{pp}=0$ and $a_{p1}=a_{p2}=\ldots =a_{pp}=0$ and also $b_p=0$. Define $\bA^*_L$ to be the $(p-1)\times (p-1)$ matrix with the last row and column in $\bA^*$ removed and define $\bb_L$ to be the $(p-1) \times 1$ vector  with the last element in $\bb$ removed.

In the following subsections we discuss special cases of model (\ref{uqdensity2}). Sampling algorithms for generating random observations from these distributions are given in Appendix \ref{samp_alg}.

\subsubsection{Truncated Gaussian distribution}

If we set $\bbeta=\boldsymbol{0}$ and assume that $\bA_L^*$ is negative definite, then we have a $(p-1)$-dimensional truncated Gaussian distribution and in this case if $\bA^*_L$ is large in magnitude and $\bmu=-\tfrac{1}{2}\bA_L^{*-1}\bb_L$ is not too close to the boundary of the simplex, then the distribution is approximately multivariate Gaussian with a general covariance matrix equal to $-\tfrac{1}{2}\bA^{*-1}_L$ and mean vector $\bmu$. This model has the same number of parameters as Aitchison's logistic normal distribution and is very flexible. 

\subsubsection{Dirichlet distribution}

If we set $\bA=\boldsymbol{0}$, then we obtain the Dirichlet distribution. 

\subsubsection{Hybrid distribution with $\bbeta$ fixed \label{hybrid}}

Often, depending on the data, there will be too many parameters to estimate in model (\ref{uqdensity2}) and some of the parameters may not be identifiable. Working with microbiome data we have found it convenient to fix the shape parameters $\beta_1,\beta_2,\ldots, \beta_p$ in advance to help control the shape of each marginal distribution near zero. For example, some of the components might be roughly Gaussian so we could preset these to $\beta_j=0$, while if the marginal distribution looks right skewed with a mode at or close to zero, then we could preset $\beta_j=-0.5$, etc. Fixing the $\beta_j$'s and estimating the other parameters $\bA^*_L$ and $\bb_L$ is analogous to choosing a Box-Cox transformation in order to reduce skewness; that is, the power parameter is chosen first and then a Gaussian distribution is estimated. Having extra shape parameters in the model can lead to difficulties in estimation and it is often convenient to treat the $\beta_j$'s as tuning constants (e.g. Scealy and Wood, 2019).  

\subsection{Discrete case: multinomial model with latent variables \label{multmodel}}

In this case we observe multivariate counts $\bx_i=(x_{i1},x_{i2}, \ldots, x_{ip})^\top$, $i=1,2,\ldots, n$, for $p$ taxa on $n$ subjects where the total count for each individual, $m_i=\sum_{j=1}^px_{ij}$, is known and assumed fixed. We assume that for $i=1,2,\ldots, n$, each $\bx_i$ given $\bu_i \in \Delta^{p-1}$ and $m_i$ are independent of each other and each have a multinomial distribution with probability mass function given by 
\begin{equation*}
f(\bx_i|\bu_i)=\frac{m_i!}{x_{1i}!x_{2i}\cdots x_{pi}!} u_{1i}^{x_{1i}}u_{2i}^{x_{2i}}\cdots u_{pi}^{x_{pi}}.
\end{equation*}
We also assume that the $\bu_i$ are iid for $i=1,2\ldots, n$, each with pdf given by (\ref{uqdensity2}). This model is fully flexible which can be seen from the marginal moment structure: 
\begin{equation}
\Var\left(\frac{x_{ij}}{m_i}\right)=\Var(u_{ij})+ \frac{\E(u_{ij})-\E (u_{ij}^2)}{m_i},
\label{moment}
\end{equation}
$ \E\left( {x_{ij}}/{m_i} \right)=\E \left( u_{ij} \right)$ and 
\begin{equation*}
\Cov \left( \frac{x_{ij}}{m_i},  \frac{x_{ik}}{m_i}\right)=\Cov \left( {u_{ij}},  {u_{ik}}\right)- \frac{\E (u_{ij} u_{ik})}{m_i}, \quad \text{for $k\neq j$},
\end{equation*}
where the expected values, variances and covariances are calculated under the model. Clearly this model allows for overdispersion and general covariance structures since model  (\ref{uqdensity2}) contains interaction terms between the components of $u_{ij}$ and $u_{ik}$.

\section{Score matching}

For estimation we apply the square root transformation
$
\bz=(\sqrt{u_1},\sqrt{u_2},\ldots, \sqrt{u_p})^\top
$
which maps the compositional data onto the positive orthant of the $(p-1)$-dimensional hypersphere defined by
\begin{equation}
S^{p-1}_+=\{ \bz=(z_1,z_2,\ldots, z_p)^{\top} \in \Bbb{R}^{p}: \quad \parallel \bz \parallel=1, \quad z_j \geq 0, \quad j=1,2,\ldots, p  \}.
\label{Splus}
\end{equation}
Note that $S^{p-1}_+$ is a $(p-1)$-dimensional Riemannian manifold with boundary and the simplex $\Delta^{p-1}$ is also  a $(p-1)$-dimensional Riemannian manifold with boundary where $\Delta^{p-1}$ is flat and $S^{p-1}_+$ is curved. 
Define $\bz^2= \left(z_1^2, z_2^2, \ldots, z_p^2  \right)^{\top}$, then on the square-root scale model (\ref{uqdensity2}) becomes  
\begin{equation}
\frac{2^p}{c_2(\bA^*,\bb,\bbeta)} \left( \prod_{j=1}^p z_j^{1+2\beta_j} \right)\exp{\left( \bz^{2\top} \bA^* \bz^2 + \bb^{\top} \bz^2   \right)}, \quad \bz =(z_1,z_2,\ldots, z_p)^{\top} \in S^{p-1}_+.
\label{qdensity2}
\end{equation}
As we will demonstrate, working on the square-root scale is more convenient for estimation than working on the simplex scale. 

\subsection{Manifolds with boundary \label{scoring}}

Mardia et al. (2016) defined a set of score matching estimators for the Bingham, Kent and other directional distributions which leads to consistent parameter estimation without the need to calculate the normalising constant. The method is similar to the score matching estimator proposed by Hyvarinen (2005, 2007), but adapted to handle data on a Riemannian manifold $M$. Note that model (\ref{qdensity2}) is also a directional distribution. However, we cannot directly apply the Mardia et al. (2016) estimators to estimate the parameters in (\ref{qdensity2}) because  our manifold   $S^{p-1}_+$ has a boundary and Theorem 1 in Mardia et al. (2016) is not valid in this case. We now modify the estimation approach proposed in Mardia et al. (2016) to handle the boundary in order to estimate the parameters in model (\ref{qdensity2}).

Let $(M,g)$ denote a connected, orientated bounded manifold with boundary $\partial M$.  From Lee (1997, p.44), Green's first identity for a manifold $M$ with boundary is given by
\begin{equation}
\int_M u \Delta v \, dV + \int_M <\nabla u, \nabla v> dV = \int_{\partial M} u N v \, d\tilde{V},
\label{green}
\end{equation}
where $\Delta$ is the Laplacian on $M$, $\nabla$ is the gradient and $<\cdot , \cdot>$ is an inner product defined on each tangent space at points $\mathbf{y} \in M$, $N$ is the outward unit normal to the boundary $\partial M$, $dV$ is the volume element of $M$
and $d\tilde{V}$ is the volume element associated with the induced metric on $\partial M$. It is also assumed that the functions $u$ and $v$ on $M$ are twice continuously differentiable. 

Next we modify Green's first identity in (\ref{green}) into a more convenient form which will be needed later for score matching.  The proof is given in Appendix \ref{th1}. 

\noindent \textbf{Theorem 1}.  \textit{Let $h: M \rightarrow \Bbb{R} $ denote a smooth function that is zero on the boundary $\partial M$. Then 
\begin{equation*}
\begin{aligned}
\int_{M} u h^2  \left( \Delta v+ 2 \frac{\langle \nabla h, \nabla v \rangle}{h} \right)  d V+ \int_M h^2 \langle \nabla  u,  \nabla v\rangle dV  & = \int_{\partial M} h^2 u  (\nabla v)^{\top}N d \tilde{V} \\
& = 0.
\end{aligned}
\end{equation*}
}

Let $f$ and $f_0$ be two probability densities on a compact Riemannian manifold $M$, defined with respect to the uniform measure on $M$, where $f$ and $f_0$ are assumed to be everywhere nonzero and twice continuously differentiable.
Next we define the following weighted Hyvarinen divergence between $f$ and $f_0$ in terms of an integrated gradient inner product for the log ratio:
\begin{align}
\Phi(f;f_0)& =\frac{1}{2}\int_M f_0 h^2 \vert \vert   \nabla \log (f)  -\nabla \log (f_0)\vert \vert^2 dV  \nonumber \\
& = \frac{1}{2}\int_M f_0 h^2 <   \nabla \log (f), \nabla \log (f)> dV  \nonumber \\
& \hskip 0.5truein + \frac{1}{2}\int_M f_0h^2  <   \nabla \log (f_0), \nabla \log (f_0)> dV \nonumber \\
& \hskip 0.5truein -  \int_M f_0h^2 <   \nabla \log (f), \nabla \log (f_0)> dV,
\label{plumb1}
\end{align}
where $h^2$ is a smooth function on $M$. We discuss the choice of weight function $h^2$ later and we assume that it is zero on the boundary  $\partial M$. Here $f$ represents the density of the model that we would like to fit to data and $f_0$ is the unknown population density. To estimate the parameters in model $f$, we need to minimise the objective function $\Phi(f;f_0)$ with respect to the parameters in $f$. In this minimisation the second integral in (\ref{plumb1}) can be ignored because it is not a function of the parameters in $f$. The third integral in (\ref{plumb1}) can be simplified using Theorem 1 and the following simple result in Mardia et al. (2016), $\partial \log{(f_0)}/\partial y_j =f_0^{-1} \partial f_0/\partial y_j$, where $y_1,y_2,\ldots, y_{p-1}$ are local coordinates assuming the manifold has dimension $p-1$. That is, 
\begin{equation*}
\begin{aligned}
 -  \int_M f_0h^2 <   \nabla \log (f), \nabla \log (f_0)> dV   & = -  \int_M f_0^{-1} f_0h^2 <   \nabla \log (f), \nabla f_0> dV  \\
& = \int_{M}  f_0 h^2  \left( \Delta \log (f)+ 2 \frac{\langle \nabla h, \nabla\log (f) \rangle}{h} \right)  d V \\
& = \int_{M}  f_0   \left( h^2 \Delta \log (f)+   {\langle \nabla h^2, \nabla\log (f) \rangle} \right)  d V. 
\end{aligned}
\end{equation*}
Hence minimising  $\Phi(f;f_0)$ over the parameters in $f$ is equivalent to minimising 
\begin{align}
\Psi(f;f_0) & = \frac{1}{2}\int_M f_0 h^2 <   \nabla \log (f), \nabla \log (f)> dV \nonumber \\
&  \hskip 0.5truein +  \int_{M}  f_0   \left( h^2 \Delta \log (f)+   {\langle \nabla h^2, \nabla\log (f) \rangle} \right)  d V.  \nonumber
\end{align}

\subsection{The sphere versus the simplex}

There are two approaches we could take for estimation. In the first we let $M=S^{p-1}_+$ and consider $S^{p-1}_+$ as being embedded in $\mathbb{R}^p$ in the standard way and switch to Cartesian coordinates $\mathbf{z}=(z_1, \ldots , z_p)^\top$ in $\mathbb{R}^p$. Following Mardia et al. (2016), the objective function becomes
\begin{align}
\tilde{\Psi}(\tilde{f};\tilde{f}_0) & = \frac{1}{2}\int_{S^{p-1}_+} \tilde{f}_0(\bz) \tilde{h}(\bz)^2 \left(  \nabla_{\bz} \log (\tilde{f}(\bz))\right)^{\top} \bP \left(  \nabla_{\bz} \log (\tilde{f}(\bz)) \right)  d S^{p-1}_+\nonumber \\
&  \hskip 0.5truein +  \int_{S^{p-1}_+}  \tilde{f}_0 (\bz)   \left( \tilde{h}(\bz)^2 \Delta \log (f)+   { \left( \nabla_{\bz} \tilde{h}(\bz)^2 \right)^{\top}\bP \nabla_{\bz}\log (\tilde{f}(\bz)) } \right)  d S^{p-1}_+,  \nonumber
\end{align}
where $S^{p-1}_+$ is defined in (\ref{Splus}), $\tilde{f}(\bz)$ is given by equation (\ref{qdensity2}), $\tilde{f}_0 (\bz)$ is the population density of $\bz$, $ \nabla_{\bz}$ is the usual Euclidean gradient, $\bP=\bI_{p}-\bz \bz^{\top}$  is a projection matrix with $i,j$th element denoted by $p_{ij}$ and 
\begin{equation}
\Delta \log (f)= \sum_{i=1}^p \sum_{j=1}^p \sum_{k=1}^p p_{ij} \partial /\partial z_{i} \left( p_{jk} \partial  \log (\tilde{f} (\bz))/\partial z_k  \right).
\label{lap}
\end{equation}

In the second approach we could avoid doing a square-root transformation and set $M=\Delta^{p-1}$ directly and use the original Cartesian coordinates $\bu$. That is, the objective function would be
\begin{align}
\tilde{\Psi}_2(\tilde{g};\tilde{g}_0) & = \frac{1}{2}\int_{\Delta^{p-1}} \tilde{g}_0(\bu) \tilde{h}(\bu)^2 \left(  \nabla_{\bu} \log (\tilde{g}(\bu))\right)^{\top} \bQ \left(  \nabla_{\bu} \log (\tilde{g}(\bu)) \right)  d \Delta^{p-1}\nonumber \\
&  \hskip 0.5truein +  \int_{\Delta^{p-1}}  \tilde{g}_0 (\bu)   \left( \tilde{h}(\bu)^2 \Delta \log (g)+   { \left( \nabla_{\bu} \tilde{h}(\bu)^2 \right)^{\top}\bQ \nabla_{\bu}\log (\tilde{g}(\bu)) } \right)  d \Delta^{p-1},  \nonumber
\end{align}
where $\tilde{g}(\bu)$ is given by equation (\ref{uqdensity2}), $\tilde{g}_0 (\bu)$ is the population density of $\bu$, $ \nabla_{\bu}$ is the usual Euclidean gradient, $\bQ=\bI_{p}-p^{-1}\boldsymbol{1}_p\boldsymbol{1}_p^{\top}$  is a fixed projection matrix which does not depend on $\bu$ due to the simplex being flat with $i,j$th element denoted by $q_{ij}$ and 
\begin{align}
\Delta \log (g) & = \sum_{i=1}^p \sum_{j=1}^p \sum_{k=1}^p q_{ij} \partial /\partial u_{i} \left( q_{jk} \partial  \log (\tilde{g} (\bu))/\partial u_k  \right) \nonumber \\
& = \sum_{i=1}^p \frac{ \partial ^2 \log(\tilde{g}(\bu))}{\partial u_i^2} - \frac{1}{p} \sum_{i=1}^p \sum_{j=1}^p \frac{ \partial ^2 \log(\tilde{g}(\bu))}{\partial u_i \partial u_j}.  \nonumber
\end{align}

It turns out that minimising $\tilde{\Psi}(\tilde{f};\tilde{f}_0)$  and minimising  $\tilde{\Psi}_2(\tilde{g};\tilde{g}_0)$ generally lead to different estimates and score matching is not invariant to transformation onto different spaces. However, they are invariant under changes of local coordinates within a space (for example, if we permute the Euclidean coordinates the estimates are eqivariant). So the question is, which objective function is best? We compared both estimators and found via simulation that minimising $\tilde{\Psi}(\tilde{f};\tilde{f}_0)$ tended to lead to more efficient  estimation than minimising $\tilde{\Psi}_2(\tilde{g};\tilde{g}_0)$. If we do a change of variable on  $\tilde{\Psi}(\tilde{f};\tilde{f}_0)$ from $\bz$ back to $\bu$ it is seen that the resulting metric is not Euclidean on the $\bu$ scale. Note that non-Euclidean metrics are commonly applied in functional data settings where the functions are densities (e.g. Srivastava et al., 2007) and when performing clustering on the simplex (e.g. Nielsen and Sun, 2019). Srivastava et al. (2007) showed that the Euclidean metric on the square-root scale is equivalent to the Fisher-Rao metric which has a connection to the Fisher Information matrix (Srivastava and Klassen, 2016, pp 114-115). This might partly explain why we get efficiency gains using a similar method. In a related context, Cressie and Read (1984) also demonstrated that their power divergence statistic in multinomial goodness of fit tests was most optimal for a transformation close to the square-root. In the rest of this paper we will focus on minimising $\tilde{\Psi}(\tilde{f};\tilde{f}_0)$.

\subsection{Choice of $h$ function}

There are various choices of $h^2$ that we could make. Following Hyvarinen (2007), the simplest choice which guarantees that $h^2$ is zero on the boundary is
\begin{equation}
 \tilde{h}(\bz)^2=\prod_{j=1}^pz_j^2. 
\label{h1}
\end{equation}
However, the issue with this function is that  it places too much weight on the interior of the simplex and will lead to a loss of efficiency. This will get worse as the dimension increases and the majority of the data is distributed close to the boundaries. 

Similar to Yu et al. (2019), we could use a capped weight function 
\begin{equation}
 \tilde{h}(\bz)^2=\min  \left(  \prod_{j=1}^pz_j^2 , a_c^2\right),
\label{h2}
\end{equation}
which reduces the weight in the middle of the simplex. The parameter $a_c$ needs to be chosen such that  $0 < a_c  < 1$. Note that $a_c$ should not be too close to zero or 1 and the optimal choice will be somewhere in between depending on the data. 

Similar to Liu et al. (2020), another choice of weight function would be the distance from the datapoint to the boundary. On the simplex scale this distance is defined in the following Lemma. The Proof is given in Appendix \ref{lem1}. 

\noindent \textbf{Lemma 1}.  \textit{Suppose $\mathbf{u}=(u_1, \ldots , u_p)^\top  \in \textrm{Int}\left (\Delta^{p-1}\right )$, the interior of $\Delta^{p-1}$,  and let $\partial \Delta^{p-1}\subset \mathbb{R}^p$ denote the boundary of $\Delta^{p-1}$.  Then}
\begin{equation}
\tilde{h}(\bu)^2=\inf_{\mathbf{v} \in \partial \Delta^{p-1}} \vert \vert \mathbf{u} -\mathbf{v}\vert \vert = C_p \min(u_1, \ldots , u_p),
\label{shortest_dist}
\end{equation}
\textit{where $C_p=\sqrt{p/(p-1)}$}

The weight function defined in (\ref{shortest_dist}) transformed to the sphere scale is then  (ignoring constants)
\begin{equation}
\tilde{h}(\bz)^2= \min(z_1^2, \ldots , z_p^2).
\label{h3}
\end{equation}
We could also put a cap in to reduce the weight in the middle of the simplex
\begin{equation}
\tilde{h}(\bz)^2= \min(z_1^2, \ldots , z_p^2,a_c^2).
\label{h4}
\end{equation}

The weight functions (\ref{h2}), (\ref{h3}) and (\ref{h4})  are not smooth. However we can still apply Theorem 1 in this case and the objective functions defined in the previous subsection are still valid. This is proved in Theorem 2 below. The proof is given in Appendix \ref{th2}.

\noindent \textbf{Theorem 2}.  \textit{Assume that $h^2$ is a piecewise smooth function on $M$ with no jump discontinuities. Then Green's first identity and Theorem 1 stated in Section \ref{scoring} are valid.}

We explore the performance of the four different weight functions in Section \ref{simstudy}.

\subsection{Estimators for the hybrid model \label{hybridest}}

In this section we simplify the score matching objective function $\tilde{\Psi}(\tilde{f};\tilde{f}_0)$ for the hybrid model defined in Section \ref{hybrid}, where $\bbeta$ is fixed and not estimated. Derivations for the Dirichlet score matching estimator (with $\bbeta$ estimated) is given in Appendix \ref{dir}.

Let $q=(p-1) + p(p-1)/2$ denote the total number of parameters in the model and let 
\begin{equation*}
\bt=(t_1,t_2,\ldots, t_q)^{\top}=(z_1^4,z_2^4,\ldots, z_{p-1}^4,2z_1^2z_2^2,2z_1^2z_3^2,\ldots,2z_{p-2}^2z_{p-1}^2,z_1^2,z_2^2,\ldots,z_{p-1}^2)^{\top}
\end{equation*}
 denote the sufficient statistics in the model and let 
\begin{equation*}
\bpi=(a_{11},a_{22},\ldots, a_{(p-1)(p-1)},a_{12},a_{13},\ldots, a_{(p-2)(p-1)},b_1,b_2,\ldots, b_{p-1})^{\top}
\end{equation*}
contain the parameters. The objective function $\tilde{\Psi}(\tilde{f};\tilde{f}_0)$ simplifies to
\begin{equation}
\Psi=\frac{1}{2} \bpi^{\top} \bW \bpi -\bpi^{\top}\left( \bd^{(1)}+ \bd^{(2)}+ \bd^{(6)} \right),
\label{objectnew}
\end{equation}
where $\bW$ is a $q \times q$ matrix with $i,j$th element denoted by $w_{ij}$ and $\bd^{(1)}$, $\bd^{(2)}$ and $\bd^{(6)}$ are vectors of dimension $q$. Each element in $\bW$ is straightforward to calculate. Define $\bmu_i =\nabla_{\bz} t_i$ and $\nu_i=\bz^{\top}\bmu_i$ for  $i=1,2,\ldots, q$. The elements of $\bmu_i$ have the form 
$4z_j^3 e_j$, $4z_jz_k^2e_j+4z_j^2z_k e_k$ or $2z_je_j$ depending on the index, where $e_j$ represents a unit vector along the $j$th coordinate axis. The elements of $\nu_i$  are of the form $4z_j^4$, $8z_j^2z_k^2$ or $2z_j^2$ depending on the index. Then $w_{ij}=\E_0 \left( \tilde{h}(\bz)^2 (\bmu_i^{\top}\bmu_j -\nu_i\nu_j)\right)$ for $i=1,2,\ldots, q$ and $j=1,2\ldots, q$, where the expectation is taken with respect to $\tilde{f}_0(\bz)$ and is estimated using sample moments.

The $i$th element of the $q$ dimensional vector $\bd^{(1)}$ is defined as $d_i=-\E_0 \left( \tilde{h}(\bz)^2 \Delta t_i\right)$, where the Laplacian can be further simplified using (\ref{lap}) for Euclidean coordinates. Firstly define $\lambda_k=k(k+p-2)$, and as noted by Mardia et al. (2016), $-\lambda_k$ is an eigenvalue of a spherical harmonic of degree $k$.  After some algebra, it follows that 
\begin{equation*}
\Delta z_j^4=-\lambda_4 z_j^4 + 12 z_j^2, \quad \Delta z_j^2z_k^2=-\lambda_4z_j^2z_k^2+2z_j^2+2z_k^2, \quad \text{and} \quad \Delta z_j^2=-\lambda_2z_j^2+2,
\end{equation*}
for $j \neq k$. 

The $\bd^{(2)}$ term depends on the specific choice of $\tilde{h}(\bz)^2$. We first derive $\bd^{(2)}$  assuming $\tilde{h}(\bz)^2$ is given by (\ref{h2}). Let
\begin{equation*}
 I_{\bz}=
\begin{cases}
 1 & \text{if $\left( \prod_{j=1}^p z_j \right) < a_c$} \\
0 & \text{otherwise.}
\end{cases}
\end{equation*}
Then after some algebra, the $q$ dimensional vector $\bd^{(2)}$ simplifies to 
$$
\bd^{(2)}=-2 \E_0 \left\{  I_{\bz}  \tilde{h}(\bz)^2 (\bd^{(3)},\bd^{(4)},\bd^{(5)})^{\top} \right\},
$$
where 
\begin{equation*}
\bd^{(3)}=(4z_1^2(1-pz_1^2), 4z_2^2(1-pz_2^2), \ldots, 4z_{p-1}^2(1-pz_{p-1}^2) ),
\end{equation*}
\begin{equation*}
\bd^{(4)}=(4z_1^2+4z_2^2-8pz_1^2z_2^2,4z_1^2+4z_3^2-8pz_1^2z_3^2, \ldots, 4z_{p-1}^2+4z_{p-2}^2-8pz_{p-1}^2z_{p-2}^2)
\end{equation*}
and 
\begin{equation*}
\bd^{(5)}=(2(1-pz_1^2),2(1-pz_2^2),\ldots, 2(1-pz_{p-1}^2)).
\end{equation*}

Alternatively, we now derive $\bd^{(2)}$  assuming $\tilde{h}(\bz)^2$ is given by (\ref{h4}). After some algebra, the $q$ dimensional vector $\bd^{(2)}$ simplifies to 
$$
\bd^{(2)}=-\E_0 \left\{  (\bd^{(3)},\bd^{(4)},\bd^{(5)})^{\top} \right\},
$$
where $\bd^{(3)}$, $\bd^{(4)}$ and $\bd^{(5)}$ are defined below.

First assume that $z_i^2=\min(z_1^2,z_2^2,\ldots, z_p^2,a_c^2)$ for some $i \in [1,2,\ldots,p]$.
Each term in $\bd^{(3)}$ represents a particular sufficient statistic $z_j^4$ for $j=1,2,\ldots,p-1$ and if $i=j$, then the $j$th term in $\bd^{(3)}$ will be equal to 
$8z_j^4(1-z_j^2)$ or if $i \neq j$ then the $j$th term will be equal to $-8z_j^4z_i^2$. 
Each term in $\bd^{(4)}$ represents a particular sufficient statistic $2z_jz_k$ for $j=1,2,\ldots, p-1$ and $k=1,2,\ldots,p-1$ where $j < k$ and if $i=j$ then the term in $\bd^{(4)}$  representing $2z_jz_k$ is equal to 
$8z_j^2z_k^2(1-z_j^2)-8z_j^4z_k^2$. If $i=k$ then the term in $\bd^{(4)}$  representing $2z_jz_k$ is equal to $8z_j^2z_k^2(1-z_k^2)-8z_j^2z_k^4$ and if both $i\neq j$ and $i \neq k$ then the term in $\bd^{(4)}$  representing $2z_jz_k$ is equal to $-16z_i^2z_j^2z_k^2$. Each term in $\bd^{(5)}$ represents a particular sufficient statistic $z_j^2$ for $j=1,2,\ldots,p-1$ and if $i=j$, then the $j$th term in $\bd^{(5)}$ will be equal to 
$4z_j^2(1-z_j^2)$ or if $i \neq j$ then the $j$th term will be equal to $-4z_j^2z_i^2$. When $a_c^2=\min(z_1^2,z_2^2,\ldots, z_p^2,a_c^2)$, then all terms within $\bd^{(3)}$, $\bd^{(4)}$ and $\bd^{(5)}$ are zero.

Finally, we derive the $\bd^{(6)}$ term. Let $\bV$ be a $q\times p$ matrix with $i,j$th element denoted by $v_{ij}$. Define $\bmu_i =\nabla_{\bz} t_i$ and $\nu_i=\bz^{\top}\bmu_i$ for  $i=1,2,\ldots, q$ the same as previously. Let $s_j=\log{(z_j)}$, for $j=1,2,\ldots,p$ and define $\bmu^{(s)}_j =\nabla_{\bz} s_j=z_j^{-1}e_j$ and $\nu_j^{(s)}=\bz^{\top}\bmu_j^{(s)}=1$ for $j=1,2,\ldots,p$. Then define
$
v_{ij}=\E_0 \left( \tilde{h}(\bz)^2 (\bmu_i^{\top}\bmu_j^{(s)} -\nu_i\nu_j^{(s)})\right)$ for $i=1,2,\ldots, q$ and $j=1,2\ldots, p$, where the expectation is taken with respect to $\tilde{f}_0(\bz)$ and is estimated using sample moments.
Next define $\bd^{(6)}=-\bV\bpi_2$, where $\bpi_2=(1+2\beta_1,1+2\beta_2,\ldots, 1+2\beta_p)^{\top}$. 

The minimum of (\ref{objectnew}) is 
\begin{equation}
\hat{\bpi}=\hat{\bW}^{-1}(\hat{\bd}^{(1)}+\hat{\bd}^{(2)}+\hat{\bd}^{(6)} )
\label{estimatescore}
\end{equation}
which is the score matching estimator for the hybrid model (the hats denote the fact we have estimated the terms using sample moments). Estimator (\ref{estimatescore}) is straightforward to calculate since it is a simple function of certain sample moments. This estimator  is applicable even when $n$ and $p$ are very large (assuming $p< n$) as long we are able to invert a large square matrix. We could also constrain $\bA_L^*$ in the model to have a special structure, e.g. diagonal with  $\bb_L=\boldsymbol{0}$, to reduce the number of estimated parameters $q$ to avoid estimation with ill-conditioned matrices. Note that it is possible to extend the estimators to the case $p>n$ by inserting a penalty term into the objective function (for example a ridge penalty), however the details of this are left for future work.

\subsection{Estimators for the multinomial model \label{multinomial}}

In this section we briefly outline how to obtain consistent estimators for the multinomial model defined in Section  \ref{multmodel}.
From properties of the multinomial distribution (Mosimann, 1962, page 67), it follows that the conditional generalised factorial moment for the model is 
\begin{equation*}
\E \left(  \frac{x_{i1}^{(\alpha_1)}x_{i2}^{(\alpha_2)}\cdots x_{i(p-1)}^{(\alpha_{p-1})}}{m_i (m_i-1)\cdots (m_i-(\sum_{j=1}^{p-1}\alpha_j) +1)}   \Big \vert \bz_i \right)=z_{i1}^{2\alpha_1}z_{i2}^{2\alpha_2}\cdots z_{i(p-1)}^{2\alpha_{p-1}},
\end{equation*}
where $x_{ij}^{(\alpha_j)}=x_{ij}(x_{ij}-1)\cdots (x_{ij}-\alpha_j+1)$ for $j=1,2,\ldots, p-1$, each $\alpha_j$ is a non-negative integer and $\bu_i=\bz_i^2=(z_{i1}^2,z_{i2}^2, \ldots, z_{ip}^2)^{\top}$. Clearly the marginal factorial moment is therefore 
\begin{equation}
\E \left(  \frac{x_{i1}^{(\alpha_1)}x_{i2}^{(\alpha_2)}\cdots x_{i(p-1)}^{(\alpha_{p-1})}}{m_i (m_i-1)\cdots (m_i-(\sum_{j=1}^{p-1}\alpha_j) +1)}  \right)=\E \left(  z_{i1}^{2\alpha_1}z_{i2}^{2\alpha_2}\cdots z_{i(p-1)}^{2\alpha_{p-1}} \right).
\label{multmom}
\end{equation}
The relationship in (\ref{multmom}) is important for estimation. The $z_{ij}$ terms are latent variables and are not observed, however the counts $x_{ij}$ are observed for the sample. Therefore we can estimate the model moments on the righthand side of (\ref{multmom}) using sample averages based on the  $x_{ij}$'s.
When $\tilde{h}(\bz)^2$ is given by (\ref{h1}), then all the moments in $\bW$ and $\bd^{(1)}$, $\bd^{(2)}$ and $\bd^{(6)}$ are even polynomial functions of $\bz$ and in this case we can consistently estimate $\bA^*_L$ and $\bb_L$ in the multinomial latent variable model using  $\hat{\bpi}=\hat{\bW}^{-1}\left( \hat{\bd}^{(1)} + \hat{\bd}^{(2)}+ \hat{\bd}^{(6)}\right)$, by plugging in the moments based on the $\bx$ data instead of $\bz$. This estimator is consistent due to the continuous mapping theorem. For this to work we further assume that a large proportion of the sample has  $m_i> 3+p$ for $i=1,2,\ldots, n$, which is usually the case for microbiome count data.  This minimum count requirement is needed because we are using higher order moments.

\subsection{Standard Errors for the hybrid model}

We now derive standard error estimates for the estimator $\hat{\bpi}=\hat{\bW}^{-1}\hat{\bd}$, where $\hat{\bd}= \hat{\bd}^{(1)} + \hat{\bd}^{(2)}+ \hat{\bd}^{(6)}$. 
The elements of $\bW$ are integrals which can be represented as 
\begin{equation}
\int_{S^{(p-1)}_+} r_{kj}(\bz) \tilde{f}_0(\bz) d \bz,
\label{int1}
\end{equation}
where the functions within $r_{kj}(\bz)$ are defined in Section \ref{hybridest} and similarly the elements of $\bd=\bd^{(1)}+\bd^{(2)}+\bd^{(6)}$ are also integrals of certain functions defined in Section \ref{hybridest}. These integrals can be estimated by sample moments, for example 
\begin{equation*}
\frac{1}{n} \sum_{i=1}^n r_{kj}(\bz_i)
\end{equation*}
is an unbiased estimator of (\ref{int1}). Now we can write
\begin{equation*}
\hat{\bW}=\frac{1}{n} \sum_{i=1}^n \bR (\bz_i),
\end{equation*}
where $\bR(\bz_i)$ is a $q \times q$ matrix function and 
\begin{equation*}
\hat{\bd}=\frac{1}{n} \sum_{i=1}^n \br (\bz_i),
\end{equation*}
where $\br (\bz_i)$ is a $q \times 1$ vector function. Define $\bW_0=\E_{0}\left( \hat{\bW}  \right)$, $\bd_0=\E_{0} \left( \hat{\bd} \right)$,  $\bpi_0=\bW_0^{-1}\bd_0$ and $\bSigma_0=\lim_{n \to \infty} n \E_0 \left\{  \left(  \hat{\bW}\bpi_0 -\hat{\bd} \right)  \left(  \hat{\bW}\bpi_0 -\hat{\bd}  \right)^{\top} \right\}$. 

Theorem 3  below describes the asymptotic behavior of the estimator $\hat{\bpi}$. The proof is very similar to Theorem 6 in Yu et. al. (2019) and is omitted.  

\noindent \textbf{Theorem 3}.  \textit{Suppose that 
\newline (C1) $\bW_0$, $\bW_0^{-1}$, $\bd_0$ and $\bSigma_0$ exist and are entry-wise finite, and
\newline (C2) there exists an $n_0$ such that, for $n \geq n_0$, $\hat{\bW}$ is a.s. invertible.
\newline Then the minimum of $\frac{1}{2} \bpi^{\top} \hat{\bW} \bpi -\bpi^{\top} \hat{\bd} $ is a.s. unique with closed form solution $\hat{\bpi}=\hat{\bW}^{-1}\hat{\bd}$. Moreover, 
\begin{equation*}
\hat{\bpi} \rightarrow_{a.s.} \bpi_0 \quad \text{and} \quad \sqrt{n} \left( \hat{\bpi}-\bpi_0 \right) \rightarrow_d N_q \left( \boldsymbol{0}, \bW_0^{-1} \bSigma_0 \bW_0^{-1}  \right) \quad \text{as} \quad n \rightarrow \infty.
\end{equation*}
}

We can estimate $\Var \left( n^{1/2} \hat{\bpi} \right)$ using $\hat{\bW}^{-1} \hat{\bSigma}_0 \hat{\bW}^{-1}$, where 
\begin{equation}
 \hat{\bSigma}_0= \frac{1}{n}\sum_{i=1}^n \left(  \bR(\bz_i)\hat{\bpi} - \br(\bz_i) \right)  \left(  \bR(\bz_i)\hat{\bpi} - \br(\bz_i)   \right)^{\top}.
\label{se}
\end{equation}

\section{Simulation \label{simstudy}}

First we summarise the estimators used in the simulation. {\it Estimator 1} is (\ref{estimatescore}) with $h$ given by (\ref{h4}), {\it Estimator 2} is (\ref{estimatescore}) with $h$ given by (\ref{h2}), {\it Estimator 3} is (\ref{estimatescore}) with $h$ given by (\ref{h1}) and {\it Estimator 4} is (\ref{estimatescore}) with $h$ given by (\ref{h3}). {\it Estimator 5} is (\ref{estimatescore}) with $h$ given by (\ref{h1}), but with the $\bu=\bz^2$ sample moments replaced by the moments of $\bx$ (see Section \ref{multinomial} and equation (\ref{multmom})). {\it SE estimator} is the standard error estimator based on (\ref{se}) of {\it Estimator 1}. {\it Estimator 6} is the moment estimator for the Dirichlet distribution, {\it Estimator 7} is the maximum likelihood estimator for the Dirichlet distribution and {\it Estimator 8} is the maximum likelihood estimator for the Dirichlet-Multinomial distribution.

We simulated $R=1000$ samples from various continuous and discrete models. Note that whenever we simulated under a discrete model we use estimated proportions $\hat{u}_{ij}=x_{ij}/m_i$ rather than the true proportions $u_{ij}$ when calculating {\it Estimators 1-4}. This is because the true proportions were latent variables and are not observed in practice. Table \ref{tab20} gives a summary of the continuous models used in the simulations and Table \ref{tab21} summarises the discrete models. The column type summarises the type of model, where tGaussian denotes truncated Gaussian distribution. The column fixed denotes which parameters are fixed and not estimated.  Table \ref{tab0} contains the choices of $a_c$ for each model (these were chosen based on examining plots of the simulated $h$ functions and choosing $a_c$ to reduce the upper tail). 

\begin{table*}[h]
\centering
{ \footnotesize
\caption{Summary of  continuous models used in simulations}
\label{tab20}
\begin{tabular}{|c|c|c|c|c|}
\hline
model & type & $p$ & parameters & fixed \\
\hline
 1& hybrid & 5 & $\bbeta=(-0.80, -0.85, 0, -0.2, 0)$, $\bb=\boldsymbol{0}$, $\bA_L^*=$ Table  \ref{tab1} & $\bbeta$, $\bb$\\
2 & hybrid & 3 &  $\bbeta=-0.75\boldsymbol{1}_3$, $\bb=\boldsymbol{0}$, $a_{11}=-63602$, $a_{12}=15145$, $a_{22}=-5694$ & $\bbeta$, $\bb$ \\
3 & tGaussian & 3 &  $\bbeta=\boldsymbol{0}$, $\bb=\boldsymbol{0}$, $a_{11}=-26.3678$, $a_{12}=5.9598$ and $a_{22}=-35.8885$.  & $\bbeta$, $\bb$ \\
4 & tGaussian & 10 & $\bbeta=\boldsymbol{0}$, $\bA_L^*=-5000\bI_9$, $\bb_L=400\boldsymbol{1}_9$ ($\bmu_L=0.04\boldsymbol{1}_9$) & $\bbeta$ \\
5 & tGaussian & 10 & $\bbeta=\boldsymbol{0}$, $\bA_L^*=-500\bI_9$, $\bb_L=40\boldsymbol{1}_9$ ($\bmu_L=0.04\boldsymbol{1}_9$) & $\bbeta$ \\
6 & tGaussian & 10 & $\bbeta=\boldsymbol{0}$, $\bA_L^*=-50\bI_9$, $\bb_L=4\boldsymbol{1}_9$ ($\bmu_L=0.04\boldsymbol{1}_9$) & $\bbeta$ \\
7 & Dirichlet & 3 &  $\bbeta=(-0.5,0.70,540)^{\top}$ & $\bA$, $\bb$ \\
8 & Dirichlet & 10 & $\bbeta=(-0.8,-0.8,-0.8,-0.8,-0.8,-0.8,-0.8,-0.8,-0.8,-0.8)$ & $\bA$, $\bb$ \\
9 & Dirichlet & 10 & $\bbeta=(9,9,9,9,9,9,9,9,9,9)^{\top}$ & $\bA$, $\bb$ \\
10 & Dirichlet & 10 & $\bbeta=(-0.8,-0.8,-0.8,-0.8,-0.8,9,9,9,9,9)$ & $\bA$, $\bb$ \\
11 & Dirichlet & 10 & $\bbeta=(-0.8,-0.8,9,9,9,9,9,9,9,9)^{\top}$ & $\bA$, $\bb$\\
12 & Dirichlet & 10 & $\bbeta=(-0.8,-0.8,-0.8,-0.8,-0.8,-0.8,-0.8,-0.8,9,9)$ &  $\bA$, $\bb$\\
\hline
\end{tabular}
}
\end{table*}

\begin{table*}[h]
\centering
{ \footnotesize
\caption{Summary of  discrete models used in simulations}
\label{tab21}
\begin{tabular}{|c|c|c|c|}
\hline
Model & type & $p$ & parameters \\
\hline
 13& hybrid-multinomial & 5 & $\bu_{i}\sim$model 1, $\bx_{i}\sim$multinomial$(\bu_i)$ \\
14 & hybrid-multinomial& 3 &  $\bu_{i}\sim$model 2, $\bx_{i}\sim$multinomial$(\bu_i)$ \\
15 & tGaussian-multinomial & 3 &  $\bu_{i}\sim$model 3, $\bx_{i}\sim$multinomial$(\bu_i)$   \\
16 & Dirichlet-multinomial & 3 & $\bu_{i}\sim$model 7, $\bx_{i}\sim$multinomial$(\bu_i)$\\
\hline
\end{tabular}
}
\end{table*}

\begin{table*}[h]
\centering
{ \footnotesize
\caption{$a_c$ choices}
\label{tab0}
\begin{tabular}{|c|c|c|c|c|c|c|}
\hline
 \multicolumn{7}{c}{{Choice of $a_c$ for $h$ given by (\ref{h4})}}  \\
\hline
\hline
model 1&  model 13 & model 14 & model 15 & model 4 & model 5 & model 6 \\
0.01 & 0.01 & 0.01 & 0.1 & 0.1 & 0.02 & 0.02 \\
model 7 & model 16 & model 8 & model 9 & model 10 & model 11 & model 12 \\ 
0.01 & 0.01 & 0.002 &0.17 & 0.002 & 0.005 & 0.001 \\
\hline
 \multicolumn{7}{c}{{Choice of $a_c$ for $h$ given by (\ref{h2})}}  \\
\hline
\hline
model 1&  model 13 & model 14 & model 15 & model 4 & model 5 & model 6 \\
0.0001 & 0.0001  & 0.001 & 0.02& 2e-07 & 1e-07  & 1e-07  \\
model 7 & model 16 & model 8 & model 9 & model 10 & model 11 & model 12 \\ 
 0.001& 0.001  &  1e-08& 6e-06 & 2e-09  & 2e-07  &  5e-11 \\
\hline
\end{tabular}
}
\end{table*}

Tables \ref{tab4}-\ref{tab12} contain the simulation results. SE denotes the true standard error of the {\it  Estimator} estimated from the $R$ simulated samples. RMSE denotes the true root mean squared error of the {\it  Estimator} estimated from the $R$ simulated samples. rbias is the relative bias of the {\it  Estimator} defined as bias/SE. $5\%$, $50\%$ and $95\%$ are the 5th, 50th and 95th percentiles respectively of the distribution of the {\it SE estimator}  of {\it Estimator 1} over the $R$ samples. We first comment on the results for Model 13 in Table \ref{tab4} which is motivated from the best fitting model for $p=5$ in Section \ref{data}.  The relative biases for {\it Estimators 1-4} are all not significant since $\mid \text{rbias} \mid << 2$. This is not surprising because $n$ is small compared to $m_i$ and the first term in (\ref{moment}) is larger than the second, and therefore for the purposes of estimation we can treat $\bu_i \approx \bx_i/m_i$.  {\it Estimator 1} and {\it Estimator 2} are the most efficient  with {\it Estimator 1} being slightly better than {\it Estimator 2}.  {\it Estimator 3} and {\it Estimator 4} performed very poorly because the $h$ distributions were highly right skewed near 0. Clearly good choices of $a_c$ made a huge difference for this model when the proportions were distributed close to the boundaries of the simplex. The {\it SE estimator} for {\it Estimator 1} performed reasonably well, it did have a tendency to underestimate the SE, but it also overestimated in a moderate number of cases. 

In Table \ref{tab4}, Model 1 is the continuous version of Model 13. In this case, {\it Estimator 1} performed better than {\it Estimator 2} and again {\it Estimator 3} and {\it Estimator 4} performed poorly. Clearly all four estimators appeared to be consistent as expected  since as $n$ increased for Model 1 all the SE's and biases reduced. The {\it SE estimator} also improved as $n$ increased. Note that the SE's are a little larger in Model 13 than they are in Model 1 when $n=92$. This is not surprising and is due to the extra multinomial variation which effectively introduces measurement error to the sample proportions. A similar increase in SE's can be seen when comparing the maximum likelihood estimators in the continuous and discrete cases for the Dirichlet distribution (compare {\it Estimator 7} with {\it Estimator 8} in Table \ref{tab11}).

\begin{table*}[h!]
\centering
{ \footnotesize
\caption{Simulation results}
\label{tab4}
\begin{tabular}{|c|c|c|c|c|c|c|c|c|c|c|c|}
\hline
&   \multicolumn{2}{c}{{\it Estimator 1}}  \vline & \multicolumn{2}{c}{{\it Estimator 2}} \vline & \multicolumn{2}{c}{{\it Estimator 3}} \vline & \multicolumn{2}{c}{{\it Estimator 4}}  \vline &  \multicolumn{3}{c}{{\it SE estimator}}  \vline \\
&SE  & rbias & SE & rbias & SE & rbias & SE & rbias & $5\%$ & $50\%$ & $95\%$\\
\hline
 \multicolumn{12}{c}{{Model 1 $n=92$}}  \\
\hline
$a_{11}$&  74004  & -0.81  &  92969  & -0.82    & 1318900   & -0.57   & 222870 & -0.85 & 25032  & 49096   & 97661  \\
$a_{22}$&  5523.1  & -0.69  &  6172.9 & -0.71    & 73205   &  -0.6  & 14120 & -0.8 & 1617.6  & 3164.7 & 6881.4   \\
$a_{33}$& 22.737  & -0.71  & 30.734  &   -0.74  & 526.07   & -0.55  & 83.171 &  -0.76  & 8.4055  & 15.9750  & 31.8920\\
$a_{44}$&  18.98  & -0.65  & 23.827 &   -0.71  & 324.17  & -0.64   & 57.995 &  -0.84   & 7.8995  & 13.8730  & 27.1250  \\
$a_{12}$&  13988  &    0.34&  16172  & 0.37   & 205000   &  0.34 & 39437 & 0.43  &  5351.7  & 9206.0  & 17042.0 \\
$a_{13}$&  1150.5 & 0.68  & 1481.5 &   0.69  & 22207  &   0.51  & 3734.3 &  0.72  & 445.41  & 804.73  & 1530.60  \\
$a_{14}$&  952.19  & -0.035  & 1163.2  & -0.055   &  11675  &  -0.07  & 2683.4 & -0.039  &  388.37  & 669.75 & 1120.10 \\
$a_{23}$&  284.29 & 0.066  &  338.59 & 0.042   & 3771.2   &  0.0083 & 705.52 & 0.014  & 107.40  & 188.49  & 327.34  \\
$a_{24}$&  270.09  &  0.53 & 309.41  & 0.54    & 3473.2   &  0.54  &  713.72 & 0.66  & 101.61 & 188.98  & 359.89   \\
$a_{34}$& 16.068  &  0.21  &   20.621 &   0.28 &  239.78  & 0.21   &  51.855 & 0.21  & 7.7004 & 12.7320  &  20.6610 \\
\hline
 \multicolumn{12}{c}{{Model 13 $n=92$, $m_i=2000$}}  \\
\hline
$a_{11}$&   95210  & -0.24  & 99687    & -0.22    &  2306000   & -0.26   &  235350 & -0.4 & 16240   & 38876   & 107860  \\
$a_{22}$& 7673.1  &  -0.59 & 7667.6   & -0.54    & 89574    & -0.49  & 15929  & -0.64 & 1407.5  & 3311.3     & 8755.1  \\
$a_{33}$&  32.796 &   -0.51  &   35.549  &  -0.48    &  895.78   &   -0.32  & 86.1 &  -0.59 & 8.140 & 18.176   & 45.422   \\
$a_{44}$& 30.341 &  -0.67  & 31.498    & -0.67    & 429.17    & -0.53  & 66.713 &  -0.78 &  8.901 & 18.377   &  43.037 \\
$a_{12}$& 17624 & -0.041  & 18209   &   -0.009   &  267490  & 0.2   & 42843  &  0.17 &  4102.0  & 8699.4   &  20312.0  \\
$a_{13}$& 1442.7  &  0.23  &  1560.6  &   0.2   &  42806   &  0.22  & 3816.9  & 0.37 &  315.69   & 711.02    &  1844.80  \\
$a_{14}$&  1156.3 &  0.27 &   1213.4  &   0.18   &  14559    &  -0.041   & 2524.2  & 0.056 & 320.58   & 668.61   &  1395.40  \\
$a_{23}$& 354.73 &  0.45 &   364.34  & 0.35    & 5231.3    &  0.061   &  788.15 &  0.2 &  96.981   & 199.720   &  412.860  \\
$a_{24}$& 377.29  &  0.47   &  382.4   &  0.42    & 4183.1    &   0.44  &  773.71 & 0.55 & 99.068  &  214.260   & 504.770   \\
$a_{34}$&22.292  &  0.096 &  23.824   & 0.18    & 320.05    &  0.19  & 50.131  &  0.22 & 8.1188   & 15.5130    & 29.3850  \\
\hline
 \multicolumn{12}{c}{{Model 1 $n=1000$}}  \\
\hline
$a_{11}$&  16120  & -0.3   &  17774  &  -0.31  & 45000  & -0.65  & 28408  & -0.46 & 11470  & 14428  & 18564  \\
$a_{22}$& 1080.4 &  -0.25  &  1131.1  &   -0.28 &  2701.2 & -0.69 & 1716.4 & -0.46  & 769.68   & 967.05  & 1296.90  \\
$a_{33}$& 4.4494  &   -0.21 & 5.3238   & -0.23    & 15.099  & -0.61  & 8.8881  & -0.4 & 3.6674  & 4.4186  & 5.5112   \\
$a_{44}$& 3.9948 & -0.22   & 4.5579  &  -0.27   & 13.03  & -0.62 & 7.5784  & -0.44  & 3.2729  & 4.0088 &  4.8840 \\
$a_{12}$& 2868.4  &   0.18  & 3096.1   &  0.19   & 8250.8  & 0.31  & 5318.8 & 0.24 & 2190.5  & 2623.4  &  3239.5  \\
$a_{13}$& 246.24  &  0.2   & 276.78  &   0.2  & 726.69  &  0.53 & 450.43 & 0.35  & 188.64  &  231.12 & 292.93   \\
$a_{14}$& 209.47 &  0.028 &  227.22  &  0.012   & 566.17  &  0.041 & 360.48  & 0.037  & 159.51  & 192.15 & 241.97   \\
$a_{23}$& 57.331  &  0.02 & 60.574   &  0.0029  & 147.51  & 0.093 & 89.45  & 0.032  & 43.992 &  53.626  & 66.327   \\
$a_{24}$& 58.272  &  0.15 &  62.323 & 0.18   & 158.28 & 0.53  & 94.859  & 0.33  & 44.395  & 55.731  & 72.201 \\
$a_{34}$& 3.6273 & 0.063  & 4.0739  &  0.092  & 10.283  & 0.07  &  6.2658 &  0.11  &  3.0215  & 3.4900  & 4.0792   \\
\hline
\end{tabular}
}
\end{table*}

We now give comment on the results in Table \ref{tab7} which examines the properties of the estimators under multinomial models for $p=3$ as $n$ and/or $m_i$ are varied. In Model 14 the underlying proportions are distributed very close to the boundaries of the simplex and the $h$ functions are highly right skewed. When $n$ is small compared with $m_i$, {\it Estimator 1} and {\it Estimator 2} were the most efficient and had small bias for Model 14, however as $n$ increased with $m_i$ fixed all of {\it Estimators 1-4} became significantly biased. {\it Estimator 5} is consistent under the multinomial model, but is very inefficient for Model 14. We need $n > 10,000$ with $m_i=2000$ fixed for {\it Estimator 5} to have better RMSE's than {\it Estimator 1}.
In Model 15 the underlying proportions were concentrated more towards the interior of the simplex making $h$ approximately symmetric.  In this case {\it Estimator 5}, {\it Estimator 4} and {\it Estimator 3} all performed similarly for Model 15 when $m_i=2000$ and all biases were not significant even when $n=10,000$. When $m_i$ was decreased to $30$, {\it Estimator 1} and {\it Estimator 2} were the best for $n=100$ and when $n=1000$ {\it Estimator 5} generally had the smallest RMSE's and performed well in this case.  In summary, {\it Estimator 5} is recommended in cases where the underlying proportions are not too concentrated at the boundaries and when $n$ is large compared with $m_i$.  When $n$ is small or similar in size to $m_i$ then {\it Estimator 1} is a better choice.

\begin{table*}[h!]
\centering
{ \footnotesize
\caption{Simulation results}
\label{tab7}
\begin{tabular}{|c|c|c|c|c|c|c|c|c|c|c|}
\hline
&   \multicolumn{2}{c}{{\it Estimator 1}}  \vline & \multicolumn{2}{c}{{\it Estimator 2}} \vline & \multicolumn{2}{c}{{\it Estimator 5}} \vline & \multicolumn{2}{c}{{\it Estimator 3}}  \vline &  \multicolumn{2}{c}{{\it Estimator 4}}  \vline \\
&RMSE   & rbias & RMSE & rbias & RMSE  & rbias & RMSE  & rbias & RMSE  & rbias \\
\hline
 \multicolumn{11}{c}{{Model 14 $n=100$, $m_i=2000$}}  \\
\hline
$a_{11}$& 40292  & -0.17  & 41156   &  -0.17  & 2403000  & 0.024  &  134460  & -0.36 & 61136  &   -0.3  \\
$a_{22}$& 4458.3   & -0.34  &   4395.8  & -0.34  & 176500   &  0.0028 &  12371 &  -0.47  &  7876.1 & -0.39   \\
$a_{12}$&  11069  & 0.32  & 10972   & 0.29  &  638920  &  -0.023 & 32567   & 0.33 &  18488  & 0.22   \\
\hline
  \multicolumn{11}{c}{{Model 14 $n=1000$, $m_i=2000$}}  \\
\hline
$a_{11}$& 15682  &   1.7  & 15727  &  1.7  &  177990  & -0.2 & 23618   &  1.3  &  22281 &  1.8    \\
$a_{22}$& 977.99   & 0.68   &  991.77   &  0.72  &  14532  &  -0.18 &  1794.4  & 0.78  & 1658.6  &  1    \\
$a_{12}$& 2777.7  &  -0.7  &  2917.8   &  -0.81  &  50449   & 0.17  &  6610.5   & -1.1 &  5981.8 &  -1.5  \\
\hline
  \multicolumn{11}{c}{{Model 14 $n=10000$, $m_i=2000$}}  \\
\hline
$a_{11}$&  15271  & 6  & 15320  &  6 &  14629  & -0.31 &   25713 & 5.2  &  23282 &  6.2  \\
$a_{22}$& 727.93  &  2.8 &   757.45  &  2.9 &  1143.3 &  -0.28  & 1769  & 3.7  & 1571.2 & 4   \\
$a_{12}$& 2082.5  &  -2.7  & 2302.2   & -3.1  & 4017.4   & 0.27 &  6572.9  &  -4.6 &  5809.3  &  -5    \\
\hline
 \multicolumn{11}{c}{{Model 15 $n=100$, $m_i=2000$}}  \\
\hline
$a_{11}$& 6.0739  & -0.21  &  5.939   & -0.21  & 8.9317   & -0.35 &  8.6041  & -0.32  & 8.7795  &  -0.3  \\
$a_{22}$& 8.3369  &  -0.2 &  8.188  &  -0.2  & 12.431  & -0.35 &  11.956  & -0.32 & 12.042  & -0.3    \\
$a_{12}$& 7.3887  & 0.12  &  7.2126  &  0.13  & 10.741  &  0.24 & 10.399    &   0.21 &  10.81  &  0.19  \\
\hline
  \multicolumn{11}{c}{{Model 15 $n=1000$, $m_i=2000$}}  \\
\hline
$a_{11}$& 1.8105  & 0.073   &  1.781   &  0.046 &  2.4059 & -0.14 &  2.3329    & -0.029 & 2.4084   & -0.022    \\
$a_{22}$&  2.4574  & 0.081  &   2.4171  & 0.053  & 3.4551  & -0.12 &   3.3523 & -0.0031 &   3.4258  & 0.0098   \\
$a_{12}$& 2.1746  & -0.1  &   2.1161 & -0.049  & 2.9926  &  0.1 &  2.914  &  0.017 & 3.0538  &  0.006  \\
\hline
  \multicolumn{11}{c}{{Model 15 $n=10000$, $m_i=2000$}}  \\
\hline
$a_{11}$& 0.58012  & 0.42   & 0.55693    & 0.35 & 0.75161  & -0.093 & 0.75779   & 0.26 &  0.78108  &  0.23  \\
$a_{22}$& 0.83089  &  0.43  &  0.8032  &  0.37 &  1.0439  & -0.09 &  1.0609  &  0.3 & 1.0963   & 0.3   \\
$a_{12}$& 0.7175  & -0.4  &  0.67564  &  -0.27 &  0.92004  & 0.083 &  0.91459   & -0.19  & 0.9686  &  -0.19  \\
\hline
 \multicolumn{11}{c}{{Model 15 $n=100$, $m_i=30$}}  \\
\hline
$a_{11}$& 8.1064   & 1.1  &  8.1064   & 1.1  & 729.39  & -0.07 & 10.041    &  1.2  &   9.856  &   1  \\
$a_{22}$& 11.923  & 1.4   &   11.923  &  1.4  & 960.01  &  -0.058 &  14.485  & 1.4  & 14.927   &  1.5   \\
$a_{12}$&  6.1635  & -0.32   &   6.1635  &  -0.32 & 856.09  & 0.061 &  9.8916    & -0.8  & 10.356   &  -0.77  \\
\hline
  \multicolumn{11}{c}{{Model 15 $n=1000$, $m_i=30$}}  \\
\hline
$a_{11}$& 7.6144  & 4.8   & 7.6144   & 4.8 & 5.7621  & -0.22 &  9.8012  &  5.6 &   9.1933  & 4.8   \\
$a_{22}$& 11.665  &  5.5  & 11.665 & 5.5  & 8.3696  & -0.23 &  14.553   & 6.2  & 15.045  &  6.3  \\
$a_{12}$&  3.5337 &  -1.8   &  3.5337  &  -1.8 & 7.3837   &  0.19 &  7.7924   &  -3.4 &  7.9285  &  -3.2   \\
\hline
  \multicolumn{11}{c}{{Model 15 $n=10000$, $m_i=30$}}  \\
\hline
$a_{11}$& 7.6345  & 16  &   7.6345  &  16  & 1.5645  & -0.066 &   9.8695  &  19  & 9.2256   &  16  \\
$a_{22}$& 11.714 &  18   & 11.714    & 18  &  2.2867  & -0.084 & 14.665   &  21  &  15.134 & 21    \\
$a_{12}$& 3.2467  &   -6.2  &  3.2467    & -6.2  & 2.0135   &  0.065 &  7.661  &   -12 &  7.7584  & -11    \\
\hline
\end{tabular}
}
\end{table*}

Table \ref{tab10} contains the simulation results for the continuous truncated Gaussian Models 4, 5 and 6  for $p=10$. Model 4 is approximately Gaussian because the truncation probability was small. In Models 5 and 6 the underlying variance was increased so that the first 9 proportions were more truncated/right skewed near zero than in Model 4. The h function (\ref{h3}) is roughly symmetric in all cases and (\ref{h1}) is roughly symmetric for Model 4, but right skewed for Models 5 and 6. In all cases rbias is not significant and {\it Estimator 1} is the most efficient, followed by {\it Estimator 2}, {\it Estimator 4} and {\it Estimator 3}. Note that {\it Estimator 3} is quite a lot worse than the other estimators in all cases even when  (\ref{h1}) is roughly symmetric because this estimator still places too much weight towards the middle of the simplex, which was worse in high dimensions since the mean was close to the boundary.

\begin{table*}[h!]
\centering
{ \footnotesize
\caption{Simulation results}
\label{tab10}
\begin{tabular}{|c|c|c|c|c|c|c|c|c|c|c|c|}
\hline
&   \multicolumn{2}{c}{{\it Estimator 1}}  \vline & \multicolumn{2}{c}{{\it Estimator 2}} \vline & \multicolumn{2}{c}{{\it Estimator 3}} \vline & \multicolumn{2}{c}{{\it Estimator 4}}  \vline &  \multicolumn{3}{c}{{\it SE estimator}}  \vline \\
&SE  & rbias & SE & rbias & SE & rbias & SE & rbias & $5\%$ & $50\%$ & $95\%$\\
\hline
  \multicolumn{12}{c}{{Model 4 $n=1000$}}  \\
\hline
$a_{11}$&  256.07  & -0.24 &  259.99  & -0.24     & 364.22    & -0.31    & 272.43 & -0.24 & 220.14   & 245.09    & 274.41  \\
$a_{22}$& 249.22   &  -0.29 &  251.88  &   -0.3  &  361.08  & -0.38   &  266.64 &  -0.3 & 220.24 & 244.51    & 279.53  \\
$a_{12}$& 165.53   & -0.036  &  169.17  &  -0.042   & 235.25    &  -0.064   & 175.02 & -0.04  & 151.80   &  164.27   &  178.11 \\
$b_1$ &  45.131  &0.12  & 47.422   &   0.12   &  74.037   & 0.2   & 47.911  &  0.12  & 38.390   & 42.257   & 46.178   \\
$b_2$ &  42.691 &  0.16 &  44.598  &  0.18   &  70.957  & 0.25   &  45.532 &  0.18  & 38.873  & 42.114    &  46.414  \\
\hline
  \multicolumn{12}{c}{{Model 5 $n=1000$}}  \\
\hline
$a_{11}$& 39.862   & -0.26  &  43.294  &  -0.27   & 120.13    & -0.79     &  52.132 & -0.33 & 33.224   & 38.115     & 45.268  \\
$a_{22}$& 39.182  &  -0.29 & 41.86  &  -0.31    & 118.23    &  -0.8   & 50.049  & -0.35 & 33.625   &  38.278    & 45.010  \\
$a_{12}$&  24.638  & 0.00014  & 26.313  & 0.0075    &  80.035  &   -0.13  & 32.977  & -0.018 &  21.892  &  23.870    & 26.277  \\
$b_1$ &  7.9481  &  0.11 & 9.347 &   0.13   &  33.692   &  0.65   &  10.813 & 0.19 & 6.8375   &    7.5284  &  8.4066 \\
$b_2$ &  7.6976 & 0.13  &  9.0989 &  0.15   & 33.017    &   0.64   & 10.275  & 0.21 & 6.8913   & 7.5222     & 8.3418  \\
\hline
  \multicolumn{12}{c}{{Model 6 $n=1000$}}  \\
\hline
$a_{11}$&  7.1114   & -0.24  &  7.2318  & -0.23     & 14.377   &   -0.27   &  9.3729 &  -0.23 & 6.0348  &  6.8321    & 7.8394  \\
$a_{22}$&  6.7013  &  -0.25 &  6.763  &  -0.25    &  14.836   &   -0.35   &   9.3549 &  -0.31 & 6.0989   &  6.8176     & 7.9081 \\
$a_{12}$&   4.6801  & -0.026 & 4.8271  &  -0.029    & 10.255    &  -0.065   &  6.4953 & -0.016 & 4.3510   &  4.7256     & 5.1556 \\
$b_1$ & 3.883  &  0.12 & 4.1051  &   0.1   & 10.128    &  0.19     & 5.2514  & 0.12 & 3.5437   & 3.8492     & 4.1898    \\
$b_2$ & 3.8312   &  0.11  &  4.0431  &  0.1   &  10.324   &  0.25   & 5.3304  & 0.18 & 3.5550   &  3.8506    &  4.1958 \\
\hline
\end{tabular}
}
\end{table*}

Table \ref{tab11} contains the simulation results for the Dirichlet distribution when $p=3$, where the first two proportions were concentrated near the zero boundary. This model was motivated by fitting the Dirichlet distribution with moment estimators to the {\it TM7} and {\it Cyanobacteria/Chloroplast} proportions from Section \ref{data} with a pooled third component.  
 The h function (\ref{h3}) is roughly symmetric and (\ref{h1}) is right skewed.  In Model 7, {\it Estimator 1}, {\it Estimator 2} and {\it Estimator 4} performed similarly, while {\it Estimator 3} was a little worse potentially due to the skewness in (\ref{h1}) and since the proportions are distributed close to zero. Interestingly,  {\it Estimators 1-4} are all more efficient than {\it Estimator 6} (moment estimator) in Model 7. However, in the discrete multinomial Model 16 case, all {\it Estimators 1-4} performed very poorly with {\it Estimator 1} the worst. Clearly it is not possible to estimate the shape parameters $\bbeta$ from discrete data with good precision using score matching estimators when some of the proportions are concentrated near zero. However, score matching worked  well for the hybrid-multinomial  and truncated Gaussian-multinomial models with $\bbeta$ fixed (see Tables \ref{tab4} and \ref{tab7}).

\begin{table*}[h!]
\centering
{ \footnotesize
\caption{Simulation results}
\label{tab11}
\begin{tabular}{|c|c|c|c|c|c|c|c|c|c|c|c|c|}
\hline
 \multicolumn{11}{c}{{Model 7 $n=92$}}  \\
\hline
&   \multicolumn{2}{c}{{\it Estimator 7}}  \vline & \multicolumn{2}{c}{{\it Estimator 6}} \vline & \multicolumn{2}{c}{{\it Estimator 3}} \vline & \multicolumn{2}{c}{{\it Estimator 2}}  \vline &  \multicolumn{2}{c}{{\it Estimator 4}}  \vline  &  \multicolumn{2}{c}{{\it Estimator 1}}  \vline\\
&RMSE   & rbias & RMSE & rbias & RMSE  & rbias & RMSE  & rbias & RMSE  & rbias  &RMSE  & rbias  \\
\hline
$\beta_1$& 0.0536   & 0.084  & 0.131     &   0.34 & 0.112  & 0.32  &  0.0902     & 0.12   &   0.0909    &  0.17    & 0.0820 & 0.09  \\
$\beta_2$& 0.207  & 0.17  &   0.562  &  0.38 & 0.392  &  0.4  &   0.289   &  0.19  &  0.293   &  0.25  & 0.308 & 0.27  \\
$\beta_3$& 70.7  &   0.18 &  173   &  0.39 &  132 &  0.44  &  82.8   &   0.22 &  106    &  0.3    &  92.1 &  0.26 \\
\hline
  \multicolumn{11}{c}{{Model 16 $n=92$, $m_i=2000$}}  \\
\hline
&   \multicolumn{2}{c}{{\it Estimator 8}}  \vline & \multicolumn{2}{c}{{\it Estimator 6}} \vline & \multicolumn{2}{c}{{\it Estimator 3}} \vline & \multicolumn{2}{c}{{\it Estimator 2}}  \vline &  \multicolumn{2}{c}{{\it Estimator 4}}  \vline  &  \multicolumn{2}{c}{{\it Estimator 1}}  \vline\\
&RMSE   & rbias & RMSE & rbias & RMSE  & rbias & RMSE  & rbias & RMSE  & rbias  &RMSE  & rbias  \\
\hline
$\beta_1$&  0.0993  & 0.18  & 0.125     & -0.72 & 0.609  & 2.3  & 1.02    &  6.5  &  1.24     &  -3  & 12.0 &  -3.8 \\
$\beta_2$& 0.306   & 0.22  &   0.452  &  -0.55 & 0.638  & 0.84  & 0.636    & 1.3  &  0.880  & -2.3    &  8.81 & -3 \\
$\beta_3$&  105  & 0.24  & 139     &   -0.58 & 180  &  0.58  &  249   &  1.7  & 370   &  -3   & 4180 & -3.7 \\
\hline
\end{tabular}
}
\end{table*}

\begin{table*}[h!]
\centering
{ \footnotesize
\caption{Simulation results}
\label{tab12}
\begin{tabular}{|c|c|c|c|c|c|c|c|c|c|c|c|c|}

\hline
&   \multicolumn{2}{c}{{\it Estimator 7}}  \vline & \multicolumn{2}{c}{{\it Estimator 6}} \vline & \multicolumn{2}{c}{{\it Estimator 3}} \vline & \multicolumn{2}{c}{{\it Estimator 2}}  \vline &  \multicolumn{2}{c}{{\it Estimator 4}}  \vline  &  \multicolumn{2}{c}{{\it Estimator 1}}  \vline\\
&RMSE   & rbias & RMSE & rbias & RMSE  & rbias & RMSE  & rbias & RMSE  & rbias  &RMSE  & rbias  \\
\hline
 \multicolumn{11}{c}{{Model 8 $n=1000$}}  \\
\hline
$\beta_1$&  0.00631& 0.089   & 0.0137  & 0.075  & 0.405  & 0.28  & 0.271 &  0.33 &  0.0279    &  -0.0045 &  0.034 & -0.0097 \\
$\beta_{10}$& 0.0063  &  0.16 & 0.0198   & 0.13 & 0.389   & 0.33 & 0.273  & 0.33   & 0.0275    &  -0.0062  & 0.0336  & 0.037  \\
\hline
 \multicolumn{11}{c}{{Model 9 $n=1000$}}  \\
\hline
$\beta_1$& 0.171  & 0.085   &  0.468  &  0.046 &  0.18  & 0.084  &  0.176 & 0.086   &   0.177   & 0.084   & 0.175  & 0.088  \\
$\beta_{10}$& 0.176  &  0.099  &  0.501   & 0.049  &  0.182  &  0.1 & 0.179  &   0.11  &   0.18   & 0.097   &  0.178 & 0.11  \\
\hline
 \multicolumn{11}{c}{{Model 10 $n=1000$}}  \\
\hline
$\beta_1$&  0.00624  & 0.12   & 0.0219  &  0.18 &  0.327  &  0.32 &  0.19  & 0.21   &   0.0191  & 0.054   & 0.0232  & 0.055  \\
$\beta_{10}$& 0.212   &   0.17  & 1.38   & 0.25  & 6.36   &  1.2 & 0.954  &  0.49  & 0.97    & 0.39   & 0.333  & 0.13  \\
\hline
 \multicolumn{11}{c}{{Model 11 $n=1000$}}  \\
\hline
$\beta_1$& 0.0063  &  0.11  &    0.021 & 0.17 &  0.0312 & 0.14  & 0.0284  & 0.031   & 0.0143     & 0.06   & 0.0163  &  0.014 \\
$\beta_{10}$& 0.193   & 0.06   &  1.31  & 0.21  & 0.933  & 0.36 & 0.329  & 0.14  & 0.59    &  0.25  & 0.248 &  0.11\\
\hline
 \multicolumn{11}{c}{{Model 12 $n=1000$}}  \\
\hline
$\beta_1$& 0.251  &  0.17  & 0.446   & 0.094  &  27  &  0.59 & 2.84  & 1.1   &  1.32    & 0.28   &  0.449 &  0.11 \\
$\beta_{10}$& 0.00643   &  0.12  & 0.0166  & 0.072 & 4.05  &  0.12 & 0.311 &  0.37   &   0.0242   &  -0.0029   &  0.0291  & -0.047  \\
\hline
\end{tabular}
}
\end{table*}

Table \ref{tab12} contains the simulation results for the Dirichlet distribution when $p=10$. In Model 8 all the proportions were highly concentrated at the boundaries of the simplex, in Model 9 all the proportions were centred in the middle of simplex and in Models 10, 11 and 12 we varied the number of proportions in the interior versus at the boundary. In general the maximum likelihood estimator was much more efficient than the other estimators except in Model 9 where {\it Estimators 1-4} were also very efficient. {\it Estimator 1} performed the best overall out of {\it Estimators 1-4}  and was often better than the moment estimator in terms of efficiency.  {\it Estimator 2} and  {\it Estimator 3} performed very poorly in Models 8, 10 and 12 when many of the proportions were concentrated near zero.

In summary for the continuous case we recommend {\it Estimator 1} in general for the hybrid and truncated Gaussian models and {\it Estimator 1} for the Dirichlet distribution if the maximum likelihood estimators are not available (this might happen if there are a small proportion of zeros in the data, but otherwise the proportions are roughly continuous). In the discrete case when $\bbeta$ is fixed (not estimated), then {\it Estimator 1} is also recommended as long as $n$ is small or of similar size relative to $m_i$. If $m_i << n$, and $n$ is large then {\it Estimator 5} is recommended instead of {\it Estimator 1} in order to reduce bias.

\section{Data application \label{data}}

We analysed a subset of the longitudinal microbiome dataset obtained from an epidemiological study carried out in a helminth-endemic area in Indonesia (Martin et al., 2018). Stool samples were collected from 150 individuals in the years 2008 (pre-treatment) and in 2010 (post-treatment). The 16s rRNA gene from the stool samples was processed and resulted in counts of 18 bacterial phyla. Whether or not the individual was infected by helminth was also determined at both timepoints (see Figure 4, Martin et al, 2018).  

Here we restricted the analysis to the year 2008 for individuals infected by helminths which resulted in a sample size of $n=94$, and we treated these individuals as being independent. Martin et al. (2018) analysed the five most prevalent phyla and pooled the remaining into one category. We analysed a different set of phyla and for demonstrative purposes, we deliberately included two with a high number of zeros.  In our analysis we have five categories (in this order): {\it TM7}, {\it Cyanobacteria/Chloroplast}, {\it Actinobacteria}, {\it Proteobacteria} and {\it pooled}. Note that {\it pooled} contains mostly {\it Firmicutes} which is the most abundant category.  The relative abundances of {\it Actinobacteria}, {\it Proteobacteria}, {\it TM7} and {\it Cyanobacteria/Chloroplast}  are $12\%$, $10\%$, $0.1\%$  and $0.3\%$ respectively. The categories {\it TM7} and {\it Cyanobacteria/Chloroplast}  contained $38\%$ and $42\%$ zeros respectively. Note that both categories {\it TM7} and {\it Cyanobacteria/Chloroplast} also contained outliers and we deleted two observations prior to the analysis resulting in a final sample size of $n=92$. 
Let $x_{ij}$, $i=1,2,\ldots, 92$ and $j=1,2,3,4,5$ represent the sample counts with total count $m_i=2000$. The sample proportions were then calculated as follows
\begin{equation*}
u_{ij}=z_{ij}^2=\frac{x_{ij}}{m_i}, \quad \text{$i=1,2,\ldots, 92$ and $j=1,2,3,4,5$}.
\end{equation*}

Prior to fitting model (\ref{uqdensity2}) to these proportions, we first performed an exploratory data analysis of the sample proportions. 
We first examined the individual marginal distributions, that is we calculated $\bv_{ij}=(u_{ij},1-u_{ij})^{\top}$, $i=1,2,\ldots, 92$ and $j=1,2,3,4,5$. Then we fitted a 2-dimensional Dirichlet distribution to each of the 5 vectors $\bv_{ij}$, $j=1,2,3,4,5$ using the moment estimator for the Dirichlet distribution. We then simulated a large sample from each of these fitted Dirichlet distributions. Denote the simulated samples by $\hat{\bv}_{ij}=(\hat{u}_{ij}, 1-\hat{u}_{ij})$, $i=1,2,\ldots, 1000000$. For the sole purpose of model diagnostics we then rounded the simulated data to the nearest integer on the count scale by calculating 
\begin{equation}
\hat{u}_{ij}^r =\text{round}\left(\hat{u}_{ij}m_i \right)/m_i,
\label{round}
\end{equation} 
in order to help mimic the discreteness in the sample data. We then compared $\hat{u}_{ij}^r$, $i=1,2,\ldots, 1000000$ to the corresponding real sample proportions $u_{ij}$, $i=1,2,\ldots, 92$ using the Kolmogorov-Smirnov test. All $5$ $p-$values were large ($> 0.25$) indicating that the Dirichlet distribution was a good fit to the individual marginal distributions. 
In comparison, the log-ratio transformation $\log{(\hat{u}_{ij}/(1-\hat{u}_{ij}))}$ was highly left skewed for {\it TM7} and {\it Cyanobacteria/Chloroplast} and the Gaussian assumption was unrealistic for these less abundant components.

\begin{figure}
\centering
\includegraphics[trim = 0cm 0cm 0cm 0cm, clip, height=12cm,width=15cm]{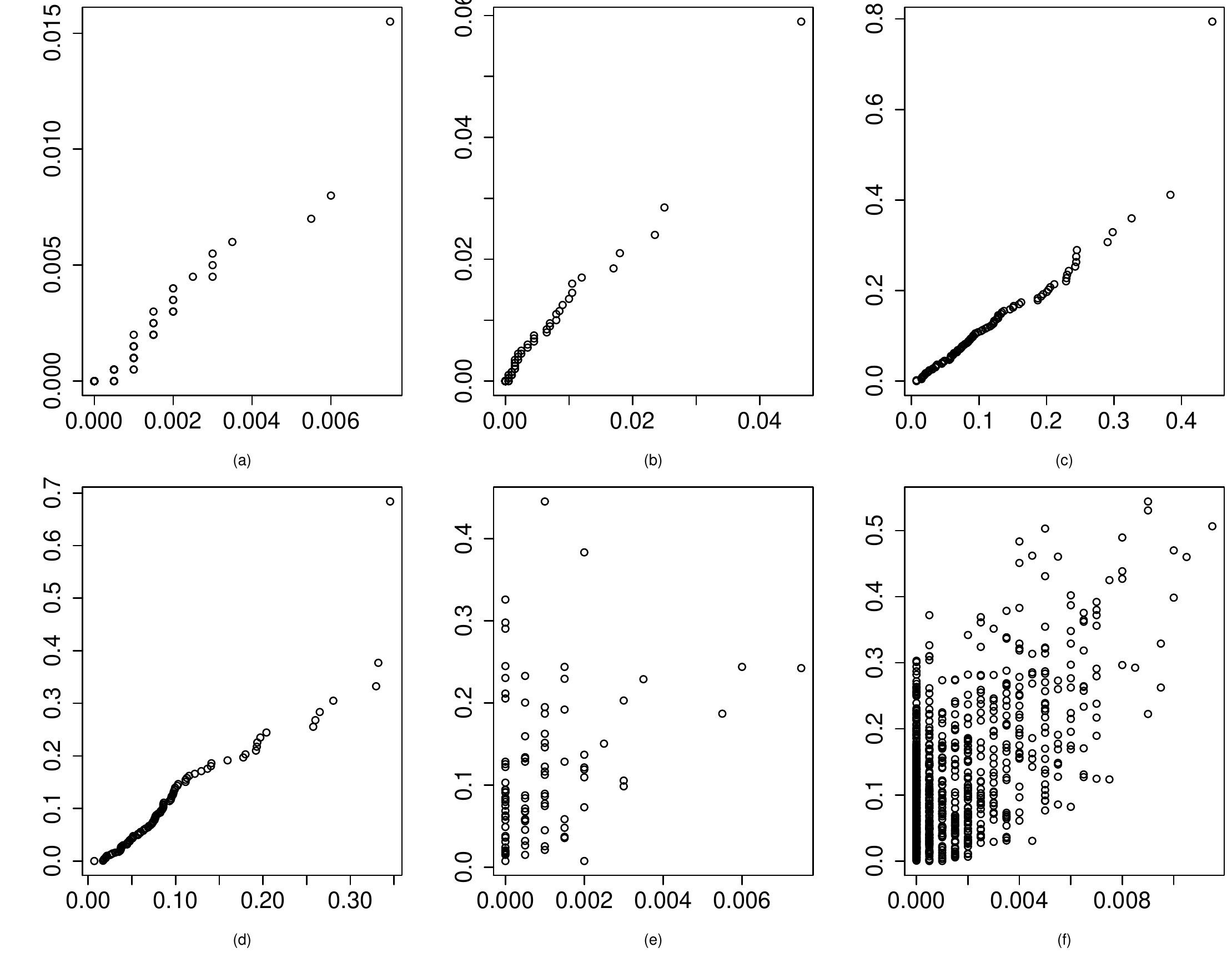}
\caption{(a)-(d): qq-plot of simulated proportions versus sample proportions for each of {\it TM7}, {\it Cyanobacteria/Chloroplast}, {\it Actinobacteria}, {\it Proteobacteria}, respectively; (e) sample  {\it Actinobacteria} proportions versus  sample  {\it TM7} proportions; (f) 1000 simulated  {\it Actinobacteria} proportions versus  the simulated  {\it TM7} proportions.}
\label{figd}
\end{figure}

Marginally, all four sample proportions $u_{ij}$, $j=1,2,\ldots, 4$ are right skewed with mode either at or close to zero with relatively high variance. We fitted the Hybrid model to the data as described in Section  \ref{hybrid} with $\bb=\boldsymbol{0}$ (this parameter is not needed because the mean and modes are close to zero relative to the variance) and the shape parameters were set to $\beta_1=-0.80$, $\beta_2=-0.85$, $\beta_3=0$, $\beta_4=-0.2$ and $\beta_5=0$.  These shape parameters were chosen to make the $p$-values large in Kolmogorov-Smirnov tests on the marginal distributions when comparing the sample data with rounded simulated data under the hybrid model. Large negative values of $\beta_j$ were needed for {\it TM7} and {\it Cyanobacteria/Chloroplast} because of the high numbers of zero's and right skewness, whereas a truncated Gaussian (or close to this $\beta_j \approx 0$) model worked reasonably well for the more abundant components. 
Table \ref{tab1} contains the parameter estimates using estimator (\ref{estimatescore}) with the $\bb$ terms omitted and we used the $h$ function given by (\ref{h4}) with $a_c=0.01$ (this was {\it Estimator 1} in the simulation study). The standard errors (SE) were calculated using (\ref{se}). All the diagonal terms in $\bA^*_L$ are significant since  $\mid\text{estimate/SE} \mid> 2$. Most of the off diagonal terms in $\bA^*_L$ are not significant, with the exception of $a_{13}$. This is not surprising because there does appear to be a positive correlation between  {\it TM7} and {\it Actinobacteria} (see Figure \ref{figd} (e) and (f)).

\begin{table*}[h]
\centering
{
\caption{Parameter estimates and standard errors for the hybrid model}
\label{tab1}
\begin{tabular}{|c|c|c||c|c|c|}
\hline
 parameter & estimate &  estimate/SE & parameter  & estimate &  estimate/SE \\
\hline
$a_{11}$ &  -127480 & -5.39  & $a_{23}$ & -8.00268     & -0.0505 \\
$a_{12}$ & 14068.4 & 1.49 & $a_{24}$ & 374.694 &   1.78  \\
$a_{13}$ & 1782.26  &  2.41 & $a_{33}$ &  -46.6387  & -2.23 \\
$a_{14}$ &  -240.077  & -0.4  & $a_{34}$ &  9.02763 & 0.758   \\
$a_{22}$ & -8191.17& -2.17 & $a_{44}$ & -39.2089  &  -2.06 \\
\hline
\end{tabular}
}
\end{table*}

For the purpose of model diagnostics, we simulated a large sample of size $n=100000$ from the hybrid model with parameters set equal to the estimates in Table \ref{tab1}. Similar to (\ref{round}) we rounded the simulated proportions to the nearest integer on the count scale. Figure  \ref{figd} (a)-(d) contains qq-plots of the simulated proportions {\it TM7}, {\it Cyanobacteria/Chloroplast}, {\it Actinobacteria}, {\it Proteobacteria} vesus the true sample proportions (compares the four sets of marginal distributions). The qq-plots are close to the $y=x$ line demonstrating that the model fits reasonably well marginally. Figure \ref{figd} (e) is a plot of the sample  {\it Actinobacteria} proportions versus the sample  {\it TM7} proportions and Figure \ref{figd} (f) is the corresponding simulated version (the first 1000 simulated values were plotted), which shows a similar pattern to the true sample ones. The hybrid model is clearly able to model positive correlations between variables, unlike the less flexible Dirichlet distribution.

\begin{table*}[h]
\centering
{
\caption{ Kolmogorov-Smirnov test results}
\label{tab2}
\begin{tabular}{|c|c|c|}
\hline
 Category &  hybrid model & Dirichlet model \\
\hline
{\it TM7} & 0.12 ($p$-value=0.15) &  0.056 ($p$-value=0.93)\\
{\it Cyanobacteria/Chloroplast} & 0.085  ($p$-value=0.52)  & 0.45 ($p$-value $<$ 2.2e-16) \\
{\it Actinobacteria} & 0.064 ($p$-value=0.84)  & 0.48  ($p$-value $<$ 2.2e-16)\\
 {\it Proteobacteria} & 0.15  ($p$-value=0.025)  & 0.44 ($p$-value = 4.4e-16) \\
{\it pooled} & 0.063  ($p$-value=0.85)  & 0.45 ($p$-value $<$ 2.2e-16) \\
\hline
\end{tabular}
}
\end{table*}

\begin{table*}[h]
\centering
{
\caption{Simulated means and standard deviations under the fitted models compared with true sample moments}
\label{tab3}
\begin{tabular}{|c|c|c|c|c|c|c|}
\hline
&   \multicolumn{3}{c}{mean}  \vline & \multicolumn{3}{c}{standard deviation}  \vline \\
Category& true & hybrid & Dirichlet & true & hybrid & Dirichlet \\
\hline
{\it TM7} & 0.0009 & 0.001 & 0.0009 & 0.001 &  0.002 & 0.001\\
{\it Cyanobacteria/Chloroplast} & 0.003 & 0.004 &  0.003 &  0.007 &  0.006 & 0.002 \\
{\it Actinobacteria} &  0.1 & 0.1 & 0.1 & 0.09 & 0.09 & 0.01\\
 {\it Proteobacteria} & 0.1 & 0.1 & 0.1 & 0.07 & 0.09 &  0.01\\
{\it pooled} & 0.8 & 0.8 &  0.8 & 0.1 & 0.1 & 0.02 \\
\hline
\end{tabular}
}
\end{table*}

For comparative purposes we also fitted a 5-dimensional Dirichlet distribution to the sample proportions based on the moment estimator and we simulated a large sample of size $n=100000$ from this fitted model and rounded them similar to (\ref{round}).  Table \ref{tab2} contains the results of Kolmogorov-Smirnov tests based on comparing the simulated proportions to the true sample proportions for each category separately. The hybrid model is clearly a very good fit since the $p$-values were generally large, with the exception of  {\it Proteobacteria} which is borderline (this is not surprising since the tail is a bit curved in Figure \ref{figd} (d)). In comparison the Dirichlet distribution is clearly a very poor fit. The Dirichlet fits the first category very well, at the expense of the four other categories where it gets the scale and shape of the marginal distributions completely wrong. Table \ref{tab3} compares the true sample moments to the simulated sample moments under both the hybrid and Dirichlet models. The hybrid model does a very good job overall of getting both the first and second order moments approximately correct. The Dirichlet model gets the means correct (this is not surprising because we used the moment estimator), but the standard deviations are grossly underestimated in most of the categories which implies that the Dirichlet distribution is getting the scale wrong. Clearly the sample proportions are over-dispersed when compared with a Dirichlet distribution.

\section{Conclusion}

In 1986 John Aitchison published a seminal book on compositional data analysis. 
Since then many articles have been written on the topic,
as this continues to be a challenging problem in the age of complex modern biological and ecological data applications. 
We proposed a new flexible truncated model for compositional data and we derived score matching estimators for the parameters in the model. 
These estimators were scalable and computationally efficient and handled many zeros in the data. The simulation study demonstrated that with an appropriate choice of weight function $h$, the score matching estimators had small bias and were generally reasonably efficient. 
Our analysis of a real microbiome dataset highlighted the deficiencies of current approaches such as the logistic normal distribution and the Dirichlet distribution for summarising microbiome data. The Dirichlet distribution was not adequate to summarise microbiome data because the data were over-dispersed. In comparison, our new hybrid model was able to accommodate proportions concentrated near zero with high right skewness. We also demonstrated that the microbiome data had an underlying {\it independence} structure because most of the off-diagonal terms in $\bA_L^*$ were not significantly different from zero.

\appendix

\section{Appendix}

\subsection{Sampling algorithms \label{samp_alg}}

\subsubsection{Truncated Gaussian Model ($\bbeta=\boldsymbol{0}$ and $\bA^*_L$ negative definite)}

Standard samplers for the multivariate Gaussian distribution can be used in this case (e.g. function mvrnorm in Library MASS in R). First let $\bmu=-\tfrac{1}{2} \bA^{*-1}_L\bb_L$ and $\bSigma=-\tfrac{1}{2}\bA^{*-1}_L$. Then 

\begin{itemize}
\item [{\it Step 1:}] generate candidate $\br=(r_1,r_2,\ldots,r_{p-1})^{\top} \backsim N_{p-1}\left(\bmu, \bSigma \right)$.
\item [{\it Step 2:}] if $\min(r_1,r_2,\ldots, r_{p-1}) \geq 0$ and $\sum_{i=1}^{p-1} r_i \leq 1$ then accept candidate $\br$, else return to Step 1.
\item [{\it Step 3:}] Let $u_1=r_1$, $u_2=r_2$, $\ldots$, $u_{p-1}=r_{p-1}$ and $u_p=1-\sum_{i=1}^{p-1}r_i$.
\end{itemize}

\subsubsection{Dirichlet distribution}

We suggest using the rdirichlet function in the R Library MCMCpack.

\subsubsection{Hybrid distribution with $\bbeta$ fixed}

In this case we propose using rejection sampling with a Dirichlet proposal. Let $\tilde{g}(\bu)$ denote density (\ref{uqdensity2}) in the general case and let  $\tilde{g}_d(\bu)$ denote density (\ref{uqdensity2}) with $\bA^{*}=\boldsymbol{0}$ and $\bb=\boldsymbol{0}$ (i.e. the Dirichlet density).  Clearly
\begin{equation*}
\begin{aligned}
\frac{\tilde{g}(\bu)}{\tilde{g}_d(\bu)} &  \propto \exp{\left( \bu^{\top} \bA^* \bu + \bb^{\top} \bu   \right)} \\
& \leq C,
\end{aligned}
\end{equation*}
where  $C$ is a constant chosen such that $$C \geq \max_{\bu \in \Delta^{p-1} }{\left( (c_2(\bA^*,\bb,\bbeta))^{-1}c_2(\boldsymbol{0},\boldsymbol{0},\bbeta) \exp{\left( \bu^{\top} \bA^* \bu + \bb^{\top} \bu   \right)} \right)}.$$ Note that a finite $C$ exists since the simplex is a bounded space. 

Then apply rejection sampling:
\begin{itemize}
\item [{\it Step 1:}] Generate candidate $\bu=(u_1,u_2,\ldots,u_p)^{\top} \backsim \text{Dirichlet} \left( \beta_1+1, \beta_2+1,\ldots, \beta_p+1  \right)$.
\item [{\it Step 2:}] Generate $r \backsim \text{uniform}(0,1)$ independent of step 1.
\item [{\it Step 3:}] Accept $\bu$ if $r \leq \exp{\left( \bu^{\top} \bA^* \bu + \bb^{\top} \bu   \right)} / C $.
\item [{\it Step 4:}] If $\bu$ is accepted, return $\bu$; otherwise go to step 1.
\end{itemize}

Note that $\bA^{*}$ may not be negative semi definite in the general case, so determining an appropriate $C$ may sometimes be difficult. We suggest guessing an initial $C$, for example $C=1$ and then update $C$ using the empirical supremum rejection sampling algorithm in Caffo et al. (2002). This algorithm worked reasonably well for the microbiome data.

\subsection{Proof of Theorem 1 \label{th1}}

For some initial background see Grigor'yan (2006) near equations (2.4) and (2.5). Green's first identity on a manifold with boundary as given at (\ref{green}) is
\begin{equation*}
\int_{M} u \Delta v d V+ \int_M \langle \nabla  u, \nabla v\rangle dV = \int_{\partial M} u N v d \tilde{V},
\end{equation*}
which is equivalent to 
\begin{equation*}
\int_{M} u \Delta v d V+ \int_M \langle \nabla  u, \nabla v\rangle dV = \int_{\partial M} u  (\nabla v)^{\top}N d \tilde{V}. 
\end{equation*}
Let $F$ be a smooth vector field defined on $M$. The divergence div $F$ is a scalar function on $M$ which is given in local coordinates by 
\begin{equation*}
\text{div} F= \frac{1}{\sqrt{\text{det}g}} \frac{\partial}{\partial y_i} \left(  \sqrt{\text{det}g} F^i \right),
\end{equation*}
where $y_1, y_2, \ldots, y_{p-1}$ are the local coordinates ($(p-1)$ is the dimension of manifold) and $g$ is the Riemannian metric tensor on $M$. By definition 
\begin{equation*}
\text{div} \nabla v = \Delta v
\end{equation*}
and therefore Green's identity becomes
\begin{equation*}
\int_{M} u \text{div} \nabla v  d V+ \int_M \langle \nabla  u, \nabla v\rangle dV = \int_{\partial M} u  (\nabla v)^{\top}N d \tilde{V}. 
\end{equation*}
Note that we can replace the term $ \nabla v$ by any smooth vector field and the identity still holds. So in the above identity replace  $ \nabla v$ with  $h^2  \nabla v$. Hence 
\begin{equation*}
\int_{M} u \text{div} h^2 \nabla v  d V+ \int_M \langle \nabla  u, h^2 \nabla v\rangle dV = \int_{\partial M} u  (h^2\nabla v)^{\top}N d \tilde{V}. 
\end{equation*}
Now note that (see Grigor'yan (2006))
\begin{equation*}
\frac{\text{div} h^2 \nabla v }{h^2} = \Delta v+ 2 \frac{\langle \nabla h, \nabla v \rangle}{h}
\end{equation*}
and therefore the identity becomes 
\begin{equation*}
\begin{aligned}
\int_{M} u h^2  \left( \Delta v+ 2 \frac{\langle \nabla h, \nabla v \rangle}{h} \right)  d V+ \int_M h^2 \langle \nabla  u,  \nabla v\rangle dV  & = \int_{\partial M} h^2 u  (\nabla v)^{\top}N d \tilde{V} \\
& = 0.
\end{aligned}
\end{equation*}
The right hand side integral is zero due to the fact that $h^2$ is zero on the boundary $\partial M$.

\subsection{Proof of Lemma 1 \label{lem1}}

 For $j=1, \ldots , p$, define face $j$ of the boundary, denoted $\mathcal{F}_j$,  by 
\[
\mathcal{F}_j=\{\mathbf{u}=(u_1, \ldots , u_p)^\top  \in \Delta^{p-1}: u_j=0\}.
\]
Then 
\[
\partial \Delta^{p-1} = \bigcup_{j=1}^p \mathcal{F}_j.
\]
The squared Euclidean distance from $\mathbf{u} \in \Delta^{p-1}$ to $\mathbf{y} \in \mathcal{F}_p$ is given by
\begin{equation}
u_p^2 +\sum_{j=1}^{p-1} (y_j - u_j)^2,
\label{squared_dist}
\end{equation}
so by adding a Lagrange multiplier term $2 \lambda (1-\sum_{i=1}^{p-1} y_i)$ and minimising over $\mathbf{y}$ we find that the optimal choice of $y_j$ is of the form $y_j=u_j+ \lambda$.   Summing from $j=1$ to $j=p-1$ and using the constraints
$\sum_{j=1}^{p-1}y_j=1$ and $\sum_{j=1}^{p-1} u_j=1-u_p$, it follows that $\lambda=u_p/(p-1)$.  Therefore, with this choice of $\lambda$, (\ref{squared_dist}) is given by
\[
u_p^2 + \sum_{j=1}^{p-1} \lambda^2= u_p^2 + (p-1) u_p^2 \frac{1}{(p-1)^2}= \frac{p}{p-1}  u_p^2.
\]
It follows from symmetry that the minimum distance from $\mathbf{u}$ to the boundary $\partial \Delta^{p-1}$ is given by (\ref{shortest_dist}).

\subsection{Proof of Theorem 2 \label{th2}}

We prove that Theorem 1 is valid for the $h^2$ function defined at (\ref{h2}). The proof for the other cases is similar. 
Break up the integrals  into separate smooth parts, for example let $M_1$ be the region of $M$ defined by  $h < a_c$ and $M_2$ be the region defined by $h > a_c$. That is
\begin{equation*}
\int_M h^2 \langle \nabla  u,  \nabla v\rangle dV  = \int_{M_1} h^2 \langle \nabla  u,  \nabla v\rangle dV+\int_{M_2} h^2 \langle \nabla  u,  \nabla v\rangle dV, 
\end{equation*}
and the first line of the equation in Theorem 1 holds on both $M_1$ and $M_2$ separately because $h^2$ is smooth within both regions. Then applying Theorem 1 
\begin{equation*}
 \int_{M_1} h^2 \langle \nabla  u,  \nabla v\rangle dV=\int_{\partial M_1} h^2 u  (\nabla v)^{\top}N d \tilde{V}  - \int_{M_1} u h^2  \left( \Delta v+ 2 \frac{\langle \nabla h, \nabla v \rangle}{h} \right)  d V
\end{equation*}
and 
\begin{equation*}
 \int_{M_2} h^2 \langle \nabla  u,  \nabla v\rangle dV=\int_{\partial M_2} h^2 u  (\nabla v)^{\top}N d \tilde{V}  - \int_{M_2} u h^2  \left( \Delta v+ 2 \frac{\langle \nabla h, \nabla v \rangle}{h} \right)  d V,
\end{equation*}
and hence
\begin{equation*}
\begin{aligned}
\int_M h^2 \langle \nabla  u,  \nabla v\rangle dV   & = \int_{\partial M_1} h^2 u  (\nabla v)^{\top}N d \tilde{V} \\
& \quad +\int_{\partial M_2} h^2 u  (\nabla v)^{\top}N d \tilde{V} - \int_{M} u h^2  \left( \Delta v+ 2 \frac{\langle \nabla h, \nabla v \rangle}{h} \right)  d V.
\end{aligned}
\end{equation*}
Note that there are two boundaries in region 1, the outer boundary $\partial M$ of the simplex and the interior boundary defined by $h=a_c$ which we denote by $\partial M_a$. Therefore 
\begin{equation}
\int_{\partial M_1} h^2 u  (\nabla v)^{\top}N d \tilde{V}=\int_{\partial M} h^2 u  (\nabla v)^{\top}N d \tilde{V}-\int_{\partial M_a} a_c^2 u  (\nabla v)^{\top}N d \tilde{V},
\label{bound1}
\end{equation}
where we are subtracting the second term in (\ref{bound1}) because this integral needs to be orientated in the reverse direction to the first integral because it is an inner boundary not an outer boundary. Also clearly 
\begin{equation}
\int_{\partial M_2} h^2 u  (\nabla v)^{\top}N d \tilde{V}=\int_{\partial M_a} a_c^2 u  (\nabla v)^{\top}N d \tilde{V}.
\label{bound2}
\end{equation}
When we add (\ref{bound1}) and (\ref{bound2}) together the $\partial M_a$ term cancels out and we obtain the same formula as Theorem 1.

\subsection{Estimation for Dirichlet distribution \label{dir}}

In this case the total number of parameters is $q=p$. Define
\begin{equation*}
\bt=(t_1,t_2,\ldots, t_p)^{\top}=(\log(z_1),\log(z_2),\ldots, \log{(z_p)})^{\top}
\end{equation*}
and the corresponding parameter vector is
\begin{equation*}
\bpi=\left(1+2\beta_1,1+2\beta_2, \ldots, 1+2\beta_p  \right)^{\top}. 
\end{equation*}
The objective function has the same form as (\ref{objectnew}), but with a different $\bW$, $\bd^{(1)}$ and $\bd^{(2)}$ ($\bd^{(6)}$ is omitted). Define $\bmu_i =\nabla_{\bz} t_i$ and $\nu_i=\bz^{\top}\bmu_i$ for  $i=1,2,\ldots, q$. The elements of $\bmu_i$ have the form $z_i^{-1} e_i$, where $e_i$ represents a unit vector along the $i$th coordinate axis and $\nu_i=1$ for all $i$. Then $w_{ij}=\E_0 \left( \tilde{h}(\bz)^2 (\bmu_i^{\top}\bmu_j -\nu_i\nu_j)\right)$ for $i=1,2,\ldots, q$ and $j=1,2\ldots, q$, where the expectation is taken with respect to $\tilde{f}_0(\bz)$ and is estimated using sample moments.

The $i$th element of the $q$ dimensional vector $\bd^{(1)}$ is defined as $d_i=-\E_0 \left( \tilde{h}(\bz)^2 \Delta t_i\right)$. After some algebra, it follows that 
\begin{equation*}
\Delta \log{(z_j)}=-\left( (p-2)+ z_j^{-2} \right), \quad j=1,2,\ldots, p.
\end{equation*}

The $\bd^{(2)}$ term depends on the specific choice of $\tilde{h}(\bz)^2$. We first derive $\bd^{(2)}$  assuming $\tilde{h}(\bz)^2$ is given by (\ref{h2}). After some algebra, $\bd^{(2)}$ simplifies to
\begin{equation*}
\bd^{(2)}=-2\E_0 \left( I_{\bz} \tilde{h}(\bz)^2 \left( (z_1^{-2}-p), (z_2^{-2}-p), \ldots, (z_p^{-2}-p)  \right)^{\top}  \right).
\end{equation*}

Alternatively, we now derive $\bd^{(2)}$  assuming $\tilde{h}(\bz)$ is given by (\ref{h4}). The term $\bd^{(2)}$ has the form $\E_0(\bd^{(3)})$, where $\bd^{(3)}$ is defined as follows.
First assume that $z_i^2=\min(z_1^2,z_2^2,\ldots, z_p^2,a_c^2)$ for some $i \in [1,2,\ldots,p]$.
Each element in $\bd^{(3)}$ represents a particular sufficient statistic $j=1,2,\ldots,p$ and if $i=j$, then the $j$th term in $\bd^{(3)}$ will be equal to 
$-2(1-z_j^2)$ or if $i \neq j$ then the $j$th term will be equal to $2z_i^2$. When $a_c^2=\min(z_1^2,z_2^2,\ldots, z_p^2,a_c^2)$, then all terms within $\bd^{(3)}$ are zero.

\section*{Acknowledgements}
The first author was supported by an Australian Research Council Discovery Early Career Researcher Award. We thank Jeanine Houwing-Duistermaat for useful conversations regarding methodology and we also thank both Jeanine and Ivonne Martin for their help with the data.


\begin{thebibliography}{4}

\bibitem[Aitchison (1986)]{ait}
\textsc{Aitchison, J.} (1986).
\textit{The Statistical Analysis of Compositional Data, Monographs on Statistics and Applied Probability, vol 25.}
Chapman \& Hall, London.


\bibitem[Bear and Billheimer (2016)]{bear}
\textsc{Bear, J. and Billheimer, D.} (2016).
A logistic normal mixture model for compositional data allowing essential zeros.
\textit{Austrian Journal of Statistics}
\textbf{45} 3--23.


\bibitem[Butler and Glasbey (2008)]{butler}
\textsc{Butler, A. and Glasbey, C.} (2008).
A latent Gaussian model for compositional data with zeros. 
\textit{Applied Statistics}
\textbf{57} 505--520; correction, \textbf{58} (2009), 141.


\bibitem[Caffo et al. (2002)]{cafo}
\textsc{Caffo, B. S., Booth, J. G. and Davison, A. C. } (2002).
Empirical supremum rejection sampling.
\textit{Biometrika}
\textbf{89} 745--754.



\bibitem[Cressie and Read (1984)]{cressie}
\textsc{Cressie, N. and Read, T. R. C.} (1984).
Multinomial goodness-of-fit tests.
\textit{Journal of the Royal Statistical Society Series B}
\textbf{46} 440--464.


\bibitem[Grigor'yan (2006)]{chen}
\textsc{Grigor'yan, A.} (2006).
Heat kernels on weighted manifolds and applications.
\textit{Contemporary Mathematics}
\textbf{398} 93--193.


\bibitem[Hyvarinen (2005)]{hy}
\textsc{Hyvarinen, A.} (2005).
Estimation of non-normalised statistical models by score matching.  
\textit{Journal of Machine Learning Research}
\textbf{6} 695--709.

\bibitem[Hyvarinen (2007)]{hy2}
\textsc{Hyvarinen, A.} (2007).
Some extensions of score matching.
\textit{Computational Statistics and Data Analysis}
\textbf{51} 2499--2512.




\bibitem[Krzysztofowicz and Reese (1993)]{kry}
\textsc{Krzysztofowicz, R. and  Reese, S.} (1993).
Stochastic bifurcation processes and distributions of fractions. 
\textit{Journal of the American Statistical Association}
\textbf{88} 345--354.



\bibitem[Lee (1997)]{lee}
\textsc{Lee, J. M.} (1997).
\textit{Riemannian Manifolds: An Introduction to Curvature}.
Springer-Verlag, New York.


\bibitem[Leininger et al. (2013)]{len}
\textsc{Leininger, T. J., Gelfand, A. E., Allen, J. M. and Silander Jr., J. A.  } (2013).
Spatial regression modeling for compositional data with many zeros.
\textit{Journal of Agricultural, Biological, and Environmental Statistics}
\textbf{18} 314--334.


\bibitem[Li (2015)]{li}
\textsc{Li, H.} (2015).
Microbiome, metagenomics, and high-dimensional compositional data analysis.
\textit{The Annual Review of Statistics and Its Application}
\textbf{2} 73--94.




\bibitem[Liu et al. (2020)]{Liu}
\textsc{Liu, S., Kanamori, T. and Williams, D. J.}  (2020).
Estimating Density Models with Truncation Boundaries.
\textit{https://arxiv.org/abs/1910.03834}.




\bibitem[Mardia (2018)]{mardia3}
\textsc{Mardia, K. V.} (2018).
A new estimation methodology for standard directional distributions. 
\textit{In: 2018 21st International Conference on Information Fusion (FUSION), IEEE, New York, 724-729}.


\bibitem[Mardia et al. (2016)]{mardia2}
\textsc{Mardia, K. V.,  Kent, J. T.  and Laha, A. K }  (2016).
Score matching estimators for directional distributions.
\textit{https://arxiv.org/abs/1604.08470}.

\bibitem[Martin et al. (2018)]{martin}
\textsc{Martin, I., Uh, H-W., Supali, T., Mitreva, M.  and Houwing-Duistermaat, J. J.}  (2018).
The mixed model for the analysis of a repeated-measurement multivariate count data.
\textit{Statistics in Medicine}  \textbf{38}  2248--2268.





\bibitem[Mosimann (1962)]{moss}
\textsc{Mosimann, J. E.}  (1962).
On the compound multinomial distribution, the multivariate $\beta$- distribution, and correlations among proportions. 
\textit{Biometrika}, \textbf{49}  65-82.



\bibitem[Nielsen and Sun (2019)]{niel}
\textsc{Nielsen, F. and Sun, K.} (2019).
Clustering in Hilbert's projective geometry: the case studies of the probability simplex and the elliptope of correlation matrices.
\textit{In: Nielsen F. (eds) Geometric Structures of Information. Signals and Communication Technology. Springer, Cham. pp. 297-331.}


\bibitem[Ongaro et al. (2020)]{Ongaro}
\textsc{Ongaro, A., Migliorati, S. and  Ascari, R.}  (2020).
A new mixture model on the simplex.
\textit{Statistics and Computing}  \textbf{30}  749--770.






\bibitem[Scealy and Welsh (2011)]{scealy3}
\textsc{Scealy, J.L. and Welsh, A. H.} (2011).
Regression for compositional data by using distributions defined on the hypersphere.
\textit{Journal of the Royal Statistical Society Series B}
\textbf{73} 351--375.


\bibitem[Scealy and Welsh (2014)]{scealy9}
\textsc{Scealy, J.L. and Welsh, A. H.} (2014).
Fitting Kent models to compositional data with small concentration.  
\textit{Statistics and Computing}
\textbf{24} 165--179.

\bibitem[Scealy and Wood (2019)]{scealy4}
\textsc{Scealy, J. L. and Wood, A. T. A.} (2019).
Scaled von Mises-Fisher distributions and regression models for paleomagnetic directional data.
\textit{Journal of the American Statistical Association}
\textbf{114} 1547--1560.









\bibitem[Srivastava et al. (2007)]{sriv2}
\textsc{Srivastava, A., Jermyn, I. and Joshi, S.} (2007).
Riemannian analysis of probability density functions with applications in vision. 
\textit{Proceedings from IEEE Conference on Computer Vision and Pattern Recognition}
\textbf{25} 1--8.


\bibitem[Srivastava and Klassen (2016)]{sriv}
\textsc{Srivastava, A. and Klassen, E. P.} (2016).
\textit{Functional and Shape Data Analysis}.
Springer-Verlag, New York.



\bibitem[Stewart and Field (2010)]{stewart}
\textsc{Stewart, C. and Field, C. A.} (2010).
Managing the essential zeros in quantitative fatty acid signature analysis.
\textit{Journal of Agricultural, Biological, and Environmental Statistics}
\textbf{16} 45--69.

\bibitem[Takasu et al. (2018)]{tak}
\textsc{Takasu, Y., Yano, K. and Komaki, F.} (2018).
Scoring rules for statistical models on spheres.  
\textit{Statistics and Probability Letters}
\textbf{138} 111--115.


\bibitem[Tsagris and Stewart (2020)]{tsagris}
\textsc{Tsagris, M. and Stewart, C.} (2020).
A folded model for compositional data analysis.
\textit{Australian and New Zealand Journal of Statistics}
\textbf{62} 249--277.


\bibitem[Yu et al. (2019)]{yu}
\textsc{Yu, S., Drton, M. and Shojaie, A.} (2019).
Generalised score matching for non-negative data.  
\textit{Journal of Machine Learning Research}
\textbf{20} 1--70.


\bibitem[Zhang and Lin (2019)]{zhang}
\textsc{ Zhang, J. and Lin, W.} (2019).
Scalable estimation and regularization for the logistic normal multinomial model.  
\textit{Biometrics}
\textbf{75} 1098--1108.



\end{thebibliography}
\end{document}